\documentclass[preprint2]{aastex}

\usepackage{breqn}
\usepackage{amsmath}
 
\shorttitle{Stochastic coupling of solar photosphere and corona}
\shortauthors{Uritsky et al.}

\begin{document}

\title{Stochastic coupling of solar photosphere and corona}

\author{Vadim M. Uritsky}
\affil{Catholic University of America at NASA Goddard Space Flight Center, Greenbelt, MD 20771 USA}
\email{vadim.uritsky@nasa.gov}

\author{Joseph M. Davila}
\affil{NASA Goddard Space Flight Center, Greenbelt, MD 20771 USA}

\author{Leon Ofman}
\affil{Catholic University of America at NASA Goddard Space Flight Center, Greenbelt, MD 20771 USA}

\author{Aaron J. Coyner}
\affil{University of Tulsa, Tulsa, OK 74104 USA}

\begin{abstract}

The observed solar activity is believed to be driven by the dissipation of nonpotential magnetic energy injected into the corona by dynamic processes in the photosphere. The enormous range of scales involved in the interaction makes it difficult to track down the photospheric origin of each coronal dissipation event, especially in the presence of complex magnetic topologies. In this paper, we propose an ensemble-based approach for testing the photosphere - corona coupling in a quiet solar region as represented by intermittent activity in SOHO MDI and STEREO EUVI image sets. For properly adjusted detection thresholds corresponding to the same degree of intermittency in the photosphere and corona, the dynamics of the two solar regions is described by the same occurrence probability distributions of energy release events but significantly different geometric properties. We derive a set of scaling relations reconciling the two groups of results and enabling statistical description of coronal dynamics based on photospheric observations. Our analysis suggests that multiscale intermittent dissipation in the corona at spatial scales $>$ 3 Mm is controlled by turbulent photospheric convection. Complex topology of the photospheric network makes this coupling essentially nonlocal and non-deterministic. Our results are in an agreement with the Parker's coupling scenario in which random photospheric shuffling generates marginally stable magnetic discontinuities at the coronal level, but they are also consistent with an impulsive wave heating involving multiscale Alfv´enic wave packets and/or MHD turbulent cascade. A back reaction on the photosphere due to coronal magnetic reconfiguration can be a contributing factor. 

\end{abstract}


\section{Introduction}
\label{sec:intro}

The Sun is an inherently multiscale system continuing to challenge theorists and experimentalists alike. It is widely believed that the observed solar activity is driven by the dissipation of free (nonpotential) magnetic energy \citep[see e.g.][and references therein]{aschwanden06, solanki06}. The free energy can be deliberated into the corona by a variety of processes in the photosphere \citep{klimchuk06}. Among the most important such processes are the injection of helicity associated with new magnetic structures \citep{falconer08, abramenko10, mcateer10, conlon10} as well as the fragmentation of the existing magnetic flux subject to magnetic footpoint shuffling due to the turbulent photospheric convection \citep{lamb08, chaouche11, abramenko05}. The amount of free coronal magnetic energy supplied through these processes is in a quasi-steady balance with the energy released via coronal heating and flaring activity \citep{simon01}, 
possibly mediated by coronal loop turbulence and waves \citep{davila87, ofman98, nigro04, buchlin05, depontieu07}.

The broad ranges of scales involved in solar activity and the presence of locally generated turbulence make it difficult to track the photospheric origin of every particular coronal event \citep[e.g., ][]{antiochos98, antiochos07, schrijver92, balke93, lawrence93, abramenko02, georgoulis05, mcateer10, parnell09, uritsky12}. Coronal brightenings occurring on strongly entangled magnetic field lines tend to interact in a nonlinear fashion producing vast regions of secondary instabilities in response to relatively small triggers \citep{klimchuk06, fuentes10}, with a noticeable fraction of magnetic discontinuities remaining silent until a local plasma instability condition is met \citep{vlahos04, aschwanden06}. Intermittent turbulent behavior of high-Reynolds number photospheric and coronal flows \citep{abramenko02, dmitruk97, berghmans98, buchlin03, nigro04, buchlin05, buchlin07, uritsky07, uritsky09}  as well as fractal diffusion in the corona \citep{aschwanden12} further increase the complexity of the coupling between the two solar regions. 

Based on this picture, the coronal response to a single reconfiguration of the photospheric field is likely to be nonlocal, non-simultaneous, and energetically disproportional to the triggering photospheric event. In many cases, searching for deterministic cause and effect links governing this interaction is close to impossible. In this paper, we adopt a different strategy focusing on statistical-physical signatures of photospheric an coronal events. We consider two types of events associated with the enhancement of the line of sight photospheric magnetic field and the extreme ultraviolet (EUV) coronal emission observed simultaneously within the same quiet solar region. The events are detected by thresholding SOHO MDI magnetograms and STEREO EUVI images, correspondingly. 

Our analysis reveals a striking similarity between ensemble-based properties of energy release events in solar photosphere and corona -- the effect we refer to as {\it stochastic coupling}. 

The results obtained show that for the detection thresholds leading to same degree of intermittency of the photospheric and coronal events, these events are described by nearly identical power-law probability distributions of the involved areas, spatiotemporal volumes, and integrated intensities. These statistical signatures indicate that temporal and spatial size of coronal heating could be 
controlled by convective shuffling motions of the photospheric magnetic field. The results are in agreement with the Parker's coupling scenario in which multiscale photospheric flows generate a large amount of marginally stable magnetic structures at the coronal level, the free magnetic energy stored by the structures being predefined by the photospheric motion \citep{parker83,parker88}. The observed photospheric dynamics should also generate a broad-band spectrum of Alfv\'{e}n waves transmitted into the chromosphere and eventually the corona \citep[e.g.,][]{davila87,ofman98, kudoh99,depontieu07}. The reported statistical constraints can help clarify physical mechanism of the photosphere - corona coupling in future studies.

The paper is organized as follows. In Section 2, we describe the data sets and the methods used in this work. Section 3 reports our main statistical results, while Section 4 is dedicated to theoretical analysis of the obtained scaling laws and their statistical interpretation. In Section 5, we discuss several plausible physical scenarios of the observed effects. Section 6 summarizes the paper.

\section{Data and methods}
\label{sec:data_methods}

\subsection{Solar images}

We studied simultaneously collected sets of SOHO MDI line of sight (LOS) magnetograms \citep{scherrer95} and STEREO-B EUVI images \citep{howard08} representing the dynamics of a quiet Sun during 17:29:00 04/05/2007 - 10:58 04/06/2007. 
The EUVI's 2048 by 2048 pixel detectors cover a field of view out to 1.7 solar radii observed in four spectral channels corresponding to the 0.1 to 20 MK temperature range. In this work, we used the 171 \r{A} bandpass corresponding to Fe IX and Fe X emission lines, with the maximum response at the solar plasma temperature $\sim 9 \times 10^5$K \citep{wulser04}.

The images were preprocessed and derotated using the standard SolarSoft software. STEREO EUVI images were then rebinned to down to the spatial resolution of SOHO MDI data (0.6 arcsec, or about 0.44 Mm) and interpolated in time using linear interpolation scheme to match the timing of the nearest MDI snapshot, with the average sampling time 66 sec. To reduce the influence of the granulation noise \citep{deforest07,lamb08} the images were smoothed in time using a 5 minute boxcar moving window. The obtained data cubes contained $770\times500\times952$ data points in the latitudinal, longitudinal and temporal directions, correspondingly. As a result of the co-alignment procedures, both data cubes (MDI and EUVI) were described by the same sets of heliographic positions and times. 

\begin{figure*}[htbp]
\begin{center}
\includegraphics*[width=7 cm]{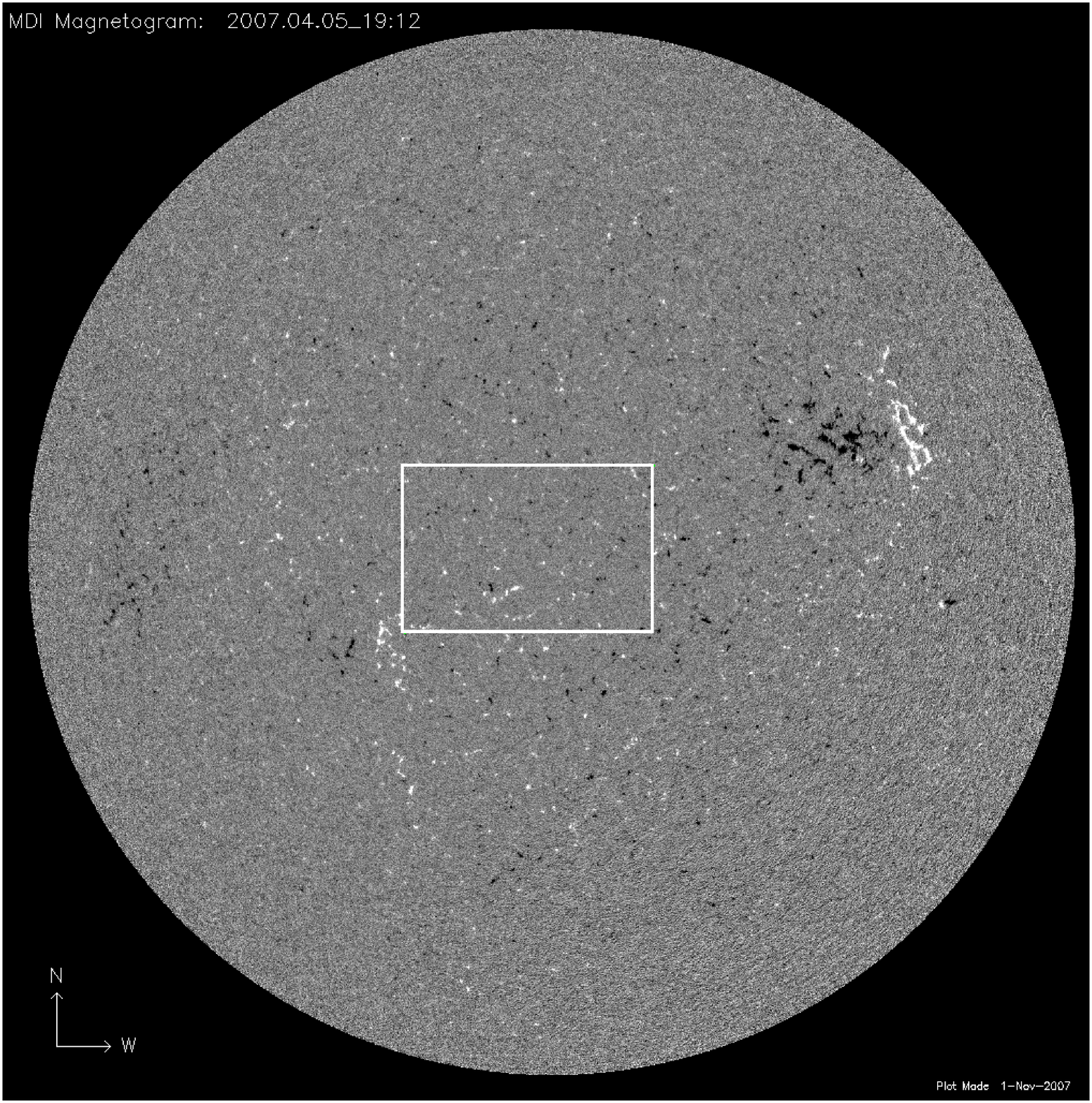} \includegraphics*[width=7.17 cm]{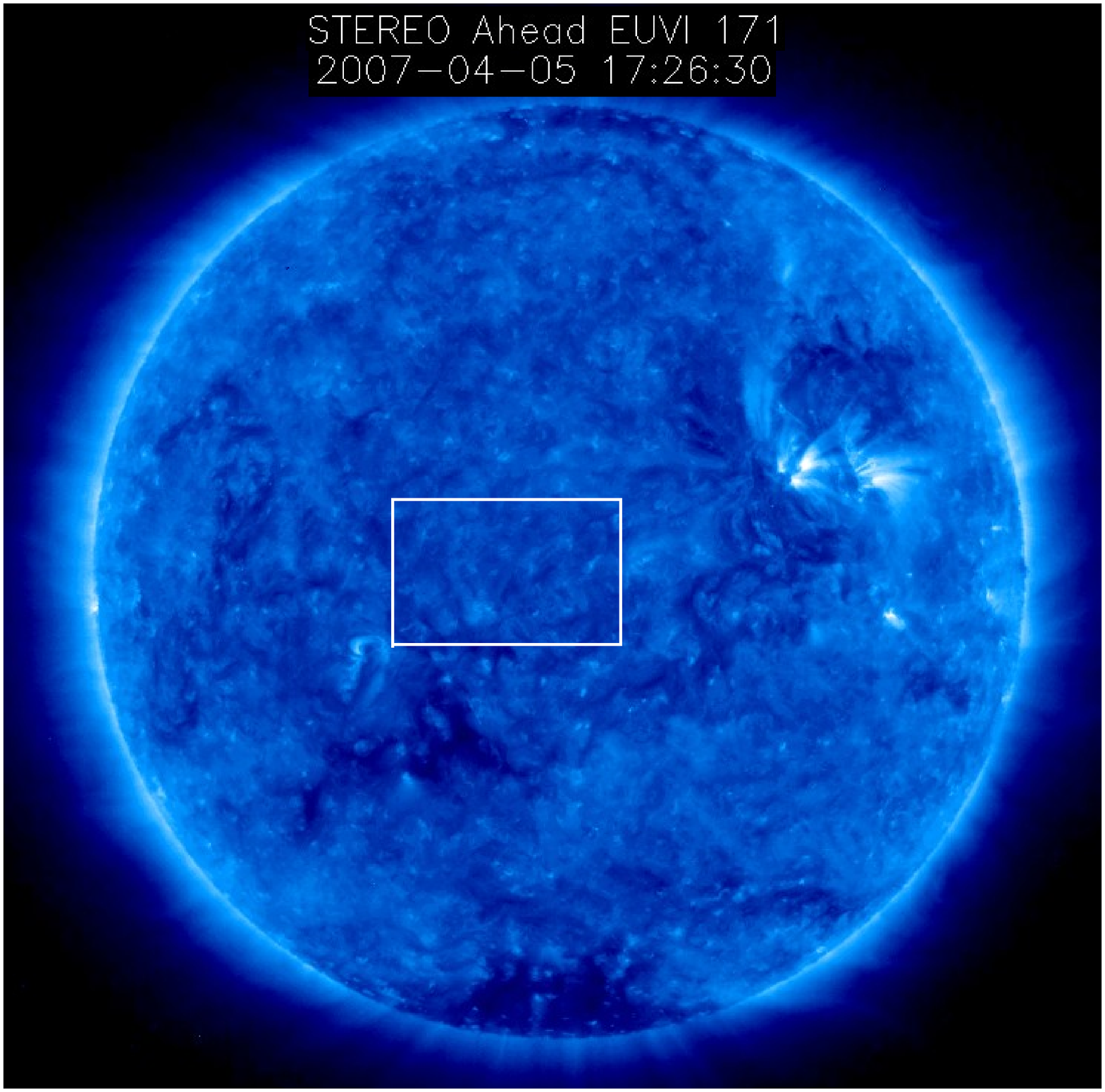} 
\caption{\label{fig1} Full-disk SOHO MDI magnetogram (left) and STEREO EUVI image (right) showing the studied quiet solar region.}
\end{center}
\end{figure*}

The overall solar activity level during the observed interval remained very low, with the GOES X-ray fluxes staying below $10^{-8}$ W m$^-2$. Fig. \ref{fig1} shows full-disk SOHO MDI and STEREO EUVI images with the studied field of view (FoV) outlined by the white frame. Two decaying active regions, NOAA 10949 and NOAA 10950, are seen outside of the FoV; neither of them produced solar flares starting from several days before till several days after the studied interval. The position of the FoV close to the disk center ensured small projection distortions. 

\begin{figure*}[htbp]
\begin{center}
\includegraphics*[width=16 cm]{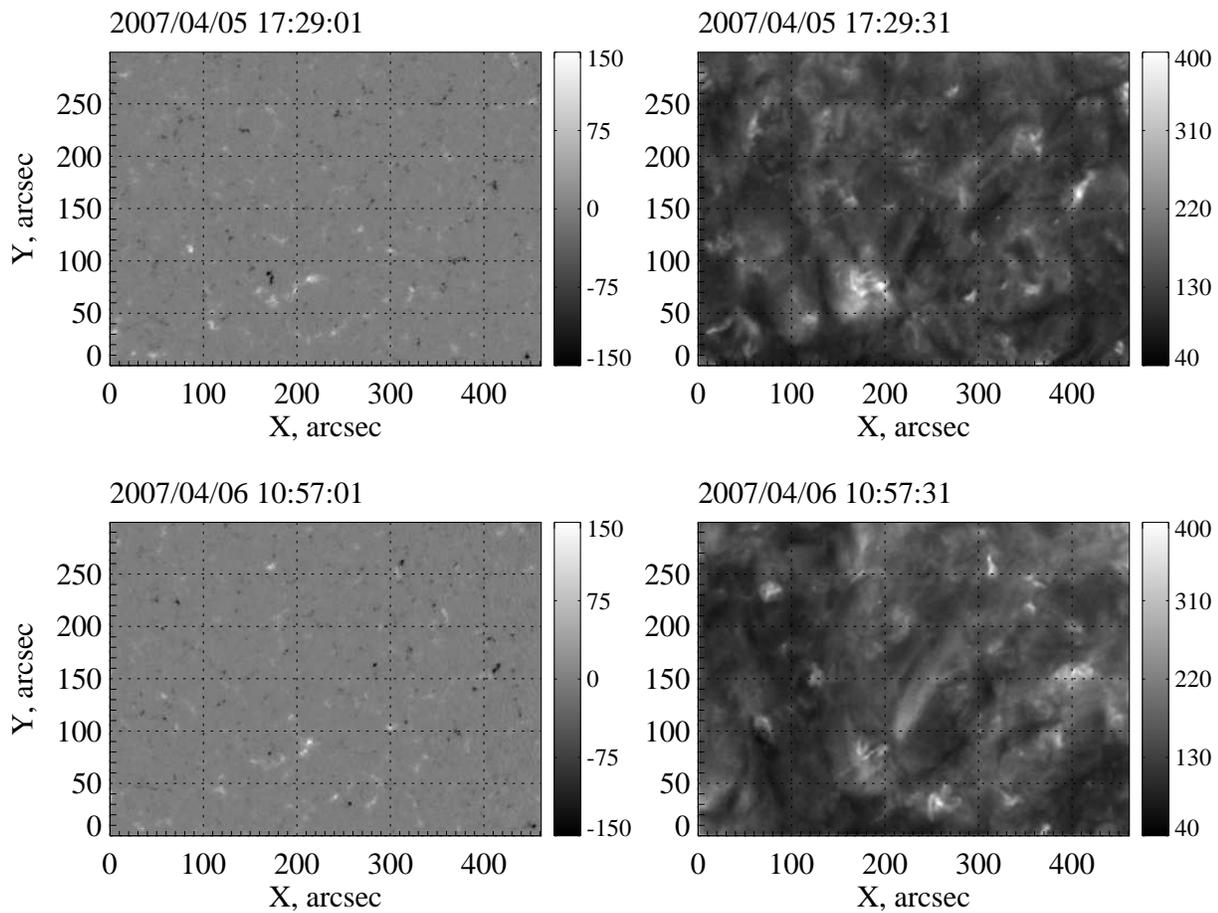}
\caption{\label{fig2} The first (top) and the last (bottom) frames of the studied conjugate sets of SOHO MDI and STEREO EUVI images (left and right columns, correspondingly).}
\end{center}
\end{figure*}

The mixed-polarity magnetic carpet and sparse coronal emission pattern found in the studied region are typical of a quiet Sun (Fig. \ref{fig2}). The LOS magnetograms reveal a hierarchy of scales reflecting meso- and supergranulation cells. The imbalance between positive and negative fluxes is about $15\%$ of the total unsigned flux in this region, with the prevalence of a negative flux. The EUVI emission field exhibits a diffuse, fractal-like distribution of intensity characteristic of a nonflaring corona. Some of the bright coronal regions coincide with magnetic field reversals, as expected  for a quiet Sun magnetic network \citep[see e.g.][]{falconer98}. The appearance of bright coronal points in the vicinity of magnetic field reversals can be a signature of low-altitude magnetic reconnection occurring on magnetic discontinuities driven by the photospheric motion \citep[see][chap. 10 for a review]{aschwanden06}.  However, the majority of coronal dissipation sites do not show obvious correlation with the underlying photospheric field.

\subsection{Event detection}

To investigate the dynamics of photospheric magnetic field and coronal emission flux, we applied a numerical method enabling the detection of spatiotemporal intermittent events in each data sets. The approach is based on a spatiotemporal tracking algorithm \citep{uritsky07, uritsky10} which identifies image features staying for more than one sampling interval above a specified detection threshold. The features are treated as subvolumes in the three-dimensional space-time, with the time history of each feature represented by a set of spatially and temporally connected pixels. 


The idea of the threshold-based spatiotemporal feature tracking has first been implemented in the context of solar magnetograms by \citet{berger96}, see also \citet{berger98}. The tracking method used in their study was based on measuring the centroid locations of the most compact and intense magnetic features. Here we invoke a different approach which does not rely on the centroid location as a guiding center. Time-evolving magnetic concentrations are treated as true 3-dimensional objects, which makes it possible to study a more general class of magnetic and emission structures whose positions cannot be adequately described by centroids. Compared to previus works, our algorithmic solution reverses the usual order of feature tracking steps (feature segmentation within an individual image followed by cross-frame feature identification within an image set, see e.g. \citet{deforest07}). Instead, we start with detecting temporal traces of features in individual solar locations, and then verify their spatial adjacency. This algorithmic solution is fast and memory-efficient, see \citet{uritsky10a} for its complete description and numerical validation.

Both MDI and EUVI data sets are characterized by broad distributions of pixel values (Fig. \ref{fig3}), encompassing more than three orders of magnitude for the magnetic field and about  two orders of magnitude for the EUVI flux. In order to make the event detection consistent with the dynamic range of each data set, we associate the detection thresholds with fixed percentile levels $p$ = 95.0, 97.0, 99.0, 99.5 $\%$ equal to the percentage of image pixels above the threshold. The dimensional thresholds corresponding to these percentiles are shown with vertical lines Fig. \ref{fig3}.  After applying the threshold, we identified all distinct path-connected sets \citep{sherbrooke96} of spatiotemporal coordinates satisfying the threshold condition and representing separate events. Table \ref{table1} provides the numerical values of the thresholds and the number of events for each percentile. Fig. \ref{fig4} shows some of the events detected in the conjugate MDI and EUVI data sets at $p = 95\%$. 

It should be noted that any threshold-based technique introduces an ambiguity with regards to the physical sizes of the detected features: the lower the threshold, the larger is the apparent size. The threshold dependence is especially strong when the probability density function of the measured parameter exhibits a slow non-exponential decay as is the case with the distributions of MDI and EUVI pixel counts (see Fig. \ref{fig3}). Since the events were selected by applying rather high intensity thresholds tuned to intermittent spikes in the studied images, the actual sizes of the events in our analysis could substantially exceed the estimated values. 

\subsection{Parameters of events and scaling ansatz}

Each of the events identified with the described technique was characterized by a set of parameters representing its spatiotemporal geometry and the flux content: 
\begin{eqnarray}\label{eq_def}
t_{1 i} &=& \underset{k \in \Lambda_i}{\text{min}} \, t_k, \,\,\,\, t_{2 i} = \underset{k \in  \Lambda_i}{\text{max}} \, t_k \\
T_i  &=& \Delta t \, (t_{2 i} - t_{1 i})  \\
A_i  &=&  (\Delta r)^2 \!\! \underset{t \in [t_{1 i}, t_{2 i}] }{\text{max}} \! \left\| \left\{  k \! \in \! \Lambda_i | t_k = t  \right\} \right\| \\
S_i  &=&  (\Delta r)^2 \!\!\!\! \bigcup_{t \in [t_{1 i}, t_{2 i}] } \!\!\!\! \left\{  k \in \Lambda_i | t_k = t  \right\} \\
l_{\{x,y\}i} &=& \underset{k, m \in \Lambda_i}{\text{max}} |\, \{x,y\}_k - \{x,y\}_m \, | \\
L_i  &=& \Delta r \, \sqrt{l_{xi}^2+l_{yi}^2} \\
V_i  &=&  (\Delta r)^2 \Delta t \, \left\| \Lambda_i  \right\| \\
E_i  &=&  (\Delta r)^2 \Delta t \, \sum_{k\in \Lambda_i}{E_k}.
\end{eqnarray}
Here, $t_{1 i}$ and $t_{2 i}$ are respectively the time of the first and the last image frame containing the $i$th event, $T_i$ is the event lifetime defined as the difference between the two, $A_i$ is the maximum instantaneous area covered by the event at any given time, $S_i$ is the area  of the spatial region involved in the event during its entire lifetime, $l_{xi}$ ($l_{yi}$) is the dimension of this region in latitudinal (longitudinal) direction, $L_i$ is the characteristic linear scale of the event, $V_i$ is the 3-dimensional spatiotemporal volume  of the event, and $E_i$ is the integrated intensity obtained by integrating the observed parameter (LOS magnetic flux or EUVI emission flux) over $V_i$. Auxiliary notations used in the above equations include the set $\Lambda_i$ of spatiotemporal coordinates (labeled with a single index) belonging to the $i$th event, discrete spatial coordinates $x$, $y$ of image pixels, discrete time step $t$, the average time step  $\Delta t$, and the linear pixel size $\Delta r$ assumed to be uniform and isotropic.

To quantify multiscale behavior of the detected photospheric and coronal events, the following scaling model has been applied:
\begin{eqnarray}
&& f(X) \propto X^{-\tau_x} \label{eq0a} \\
&& X(L) \equiv \text{E}[X|L] \propto L^{D_x} \label{eq0b}
\end{eqnarray}
in which $X \in \left\{ T, L, A, S, V, E \right\}$ is one of the event parameters, $f(X)$ is its probability density distribution, $\tau_x$ is the distribution exponent of $X$ describing the log-log decay of $f(X)$, and $D_x$ is the geometric exponent characterizing the power-law scaling of the conditional expectation value of the parameter $X$ with respect to the linear size $L$ \citep{biham01}. The scaling ansatz defined by eq. \eqref{eq0a}-\eqref{eq0b} is widely used in the theory of nonequilibrium critical phenomena, in particular, for describing processes of critical percolation \citep{munoz99}, fractal surface growth \citep{barabasi95}, and self-organized criticality \citep{bak87, bak88}. 

The power-law scaling model \eqref{eq0a} - \eqref{eq0b} was applied to the data within limited ranges of scales as explained below. Throughout this paper, we use the ``hat'' symbol to distinguish coronal scaling exponents $\hat{\tau}_x$  and $\hat{D}_x$ from their photospheric counterparts $\tau_x$ and $D_x$.




\section{Statistical results}

\subsection{Geometric and probabilistic scaling}
\label{sec:results}

Fig. \ref{fig5} shows the statistical dependence of magnetic and emission event parameters (lifetime $T$, peak area $A$, total area $S$, spatiotemporal volume $V$ and integrated flux $E$) on the linear size $L$ of the event obtained for the $ p = 95 \%$ threshold. Overplotted with the scatterplots are averaged regression trends (solid lines)
computed by coarse-graining the data using exponentially increasing $L$-bins, with error bars corresponding to three standard errors of the mean within each bin. 

The parameters plotted in Fig. \ref{fig5} grow with $L$ as approximate power-laws in the intermediate ranges of scales where the regression trends are close to straight lines on the log-log scale. The power law behavior is destroyed at the smallest and largest scales. This effect is commonly observed in real-life complex dissipative system for a multitude of reasons. The restricted scaling ranges may represent an insufficient resolution of the data, physical finite-size effects and poor statistics \citep{robinson94, uritsky06}, to name a few. To obtain reliable scaling exponents, it is important to understand what exactly causes the distortions, and focus on the intermediate scales which are free of them. 


The range of power-law scaling is particularly narrow for the geometric scaling of the event lifetime $T$ as well as the parameters $V$ and $E$ involving time-domain integration, suggesting that the accuracy of temporal measurements could be the underlying cause of the observed scaling distortion. Our analysis shows that this is indeed the case. To demonstrate this effect, we constructed a probabilistic model of the lifetime scattering subject to observational constraints imposed by the smallest duration of the events $T_{min} = 3 \, \Delta t \sim 200$ s included in the statistics, and the largest duration $T_{max} = 952\, \Delta t \sim 6.3 \times 10^4$ s dictated by the length of the observational interval. The conditional probability density $f(T|L)$ was simulated by the widely used lognormal model

\begin{eqnarray}\label{eq_lognorm}
f_L(T) & \!\! = & \frac{1}{T \nu_T \sqrt{2 \pi}} \text{e}^{ -\frac{ (\text{ln}(T) - \mu_T)^2 } {2 \nu_T^2} } \\
\nu_T  & = & \text{ln} \left( 1 + \frac{\text{Var}(T)}{(\text{E}(T))^2} \right) \\
\mu_T  & = & \text{ln}\left( \text{E}(T)\right) - \frac{1}{2}\nu_T^2
\end{eqnarray}
in which the expected value obeys the exact power law dependence $\text{E}(T) = a_T L ^{D_T}$ parametrized by the data, and the variance $\text{Var}(T)$ scales quadratically with $\text{E}(T)$. We recruit the lognormal distribution as the simplest heavy-tailed distribution supported on semi-infinite interval and possessing finite statistical moments. The analysis below can be reproduced with any other relevant distribution function.

The limited range of the observed $T$ values leads to an error in the determination of the characteristic lifetime at a given $L$ which can be approximated by the truncated mean value

\begin{eqnarray}\label{eq_avrmodel}
\left\langle T \right\rangle_L = \frac{ \int\limits_{T_{min}}^{T_{max}} \!\!\! T f_L(T) dT }{\int\limits_{T_{min}}^{T_{max}} \!\!\! f_L(T) dT}.
\end{eqnarray}
The truncated mean approaches the theoretical first moment in the double limit $T_{min} \rightarrow 0$, $T_{max} \rightarrow \infty$. For a finite lower limit, some of the short-living events are not resolved resulting in an overestimation of the characteristic lifetime at small spatial scales. Similarly, a finite $T_{max}$ truncates some of the long-living events yielding an underestimated $T$ value at the largest scales. 

The described effects are evident in Fig. \ref{fig5_1} which presents the linear scale dependence of the truncated mean $\left\langle T \right \rangle_L$ (solid curve) on the theoretical scaling (dashed diagonal line). The plots are constructed using the fitting parameters $a_T = 88.3$ and $D_T = 2.02$ for SOHO MDI data processed at the 95 $\%$ percentile level. The least-square fits refer to the range of linear scales shown with vertical dash-dotted lines. Due to significant statistical scattering of the lifetimes, the departure of $T(L)$ from the power law takes place well above $T_{min}$ and also well below $T_{max}$. The probability density plots shown on the bottom of Fig. \ref{fig5_1} visualize the observable ranges of $T$ values (shaded areas under the probability curves) for small-, intermediate-, and large-scale events.


The interval of linear sizes bounded by the values $L_1$=3 Mm and $L_2$=12 Mm shows a robust $T(L)$ power-law scaling. This scaling interval was used in all further calculations involving both MDI and EUVI data to ensure consistent statistical estimates. 

The tilted dashed lines in each of the panels of Fig. \ref{fig5} show log-log slopes of the regression fits at $L \in (L_1,L_2)$. The studied ranges of $L$ scales are shown in each of the Fig. \ref{fig5} panels by vertical dot-dashed lines. The slopes provide estimates of the geometric exponents $D_x$ and $\hat{D}_x$. 
The triangle symbols added to the $T(L)$ panel of SOHO MDI data is the truncated mean model (\ref{eq_avrmodel}) which provides a reasonable fit to the distorted scaling behavior.


Probability distributions of MDI and EUVI events for four different percentile thresholds are presented in Fig. \ref{fig6}. The power law slopes of the distributions yielding the exponents $\tau_x$ and $\hat{\tau}_x$ were measured within the ranges of scales that are consistent with the chosen range of linear sizes: 
\begin{eqnarray}
X \in \left( a_x L_1 ^{D_x}, a_x L_2 ^{D_x} \right)  \ \ \mbox{ for the photosphere} \label{eq1} \\
\hat{X} \in \left( \hat{a}_x L_1 ^{\hat{D}_x}, \hat{a}_x L_2 ^{\hat{D}_x} \right)  \ \ \mbox{ for the corona} \label{eq2},
\end{eqnarray}
where $D_x$ and $\hat{D}_x$ are the geometric exponents and $a_x$ and $\hat{a}_x$ are the regression coefficients evaluated for the two solar regions. The matching intervals of scales computed based on eq. \eqref{eq1}-\eqref{eq2} are shown by horizontal dash-dotted vertical lines in Fig. \ref{fig5} and vertical lines in Fig. \ref{fig6}. It can be seen that the distributions decay algebraically, consistent with earlier observations of power law  statistics in solar photosphere and corona \citep{crosby93, abramenko02, aschwanden02, aschwanden11, crosby11, aschwanden13}.

The broad-band power-law behavior of MDI distributions suggests that the dynamics of the LOS photospheric magnetic field is an inherently multiscale process \citep{uritsky12}. Our threshold-based definition of magnetic events is clearly not limited to the injection of new magnetic bipoles into the photosphere and their subsequent submergence into the convection zone. In essence, we deal with an apparent dynamics associated with crossings of the percentile detection threshold by the local magnetic field. The resulting sets of events likely contain a combination of truly emerging and submerging magnetic bipoles as well as magnetic flux coalescence, fragmentation, cancellation, injection of upwardly propagating Alfv\'{e}nic wave packets \citep{depontieu07}, and other effects giving rise to spatiotemporal variability of the photospheric field. In Section \ref{sec:scaling_relations}, we provide evidence that multiscale statistics of the photospheric events could be caused by turbulent convective flows (see e.g. \citet{lamb08, chaouche11, crouch07, simon95, simon01}).

In their turn, the values of EUVI distribution exponents suggest that solar corona operates in a highly intermittent turbulent state consistent with models of self-organized criticality or intermittent turbulence \citep{bak87, lu91, uritsky07}. It should be emphasized that the exponents reported here are obtained using spatiotemporal definition of events which is conceptually much closer to measuring avalanches in numerical simulations of self-organized criticality than most of the definitions used in previous works on flare statistics by other authors —- except for a few case studies \citep{berghmans98, berghmans99} focusing on specific coronal conditions. The exponent $\hat{\tau}_E <2$ indicates that the coronal dynamics is likely to be dominated by large events as opposed to Parker's scenario of nanoflare heating \citep{parker88}. Energy distribution exponents below 2 have been obtained in many previous studies \citep[e.g.][]{aschwanden00a, aschwanden00b}. Such exponent values suggest that small heating events are not frequent enough to account for the bulk coronal heating. The difficulty with ascribing this role to the large events is that they are unable to supply the required energy loss rates of $10^5-10^7$ erg/cm$^2$/s \citep[see e.g.][]{athay66}. 

Table \ref{table2} provides a summary of the geometric and distribution scaling exponents in the photosphere and corona for all four detection thresholds. The table also contains auxiliary scaling indices $\Delta_x \equiv D_x(\tau_x -1)$ and $\hat{\Delta}_x \equiv \hat{D}_x(\hat{\tau}_x -1)$. For a given threshold and data type, these indices should be independent of $X$ (constant across all event parameters) due to the probability conservation implying $f(X)dX \sim f(L)dL$ for any $X$. A departure of $\Delta_x$ ($\hat{\Delta}_x$ ) from the predicted values $D_L(\tau_L - 1) \equiv \tau_L - 1$ ($\hat{D}_L(\hat{\tau}_L - 1) \equiv \hat{\tau}_L - 1$) by more than $\sim 10 \%$ signals an insufficient statistics. Such cases are marked by asterisk.

Table \ref{table2} shows that the probability distributions of MDI and EUVI events corresponding to the same percentile level are characterized by significantly different power law slopes yielding distinct distribution exponents. For instance, most of the $\tau_x$ values describing the photospheric events at the $p=95 \%$ threshold are substantially higher compared to the corresponding $\hat{\tau}_x$ values describing EUVI events obtained at this threshold. 

We found that the difference in the scaling behavior of the distribution functions of the two solar regions originates in the difference of the photospheric and coronal intermittency observed at the same percentile threshold. Table \ref{table1} shows that for any given percentile threshold, the number of MDI events is considerably larger compared to the number of EUVI events. For the lowest applied threshold, the discrepancy reaches an order of magnitude. This suggests that for the same $p$, the magnetic events are much more intermittent than the coronal events. Applying a smaller threshold to a spiky MDI magnetogram results in much more threshold crossings, and hence more events. Due to a smoother spatial pattern of the coronal emission field, the same procedure leads to a less dramatic increase in the number of the detected coronal events. 

Physically, it is clear that comparing populations of MDI and EUVI events obtained at the same detection threshold is not justified because such events represent different types of conditions characterized by incomparable occurrence rates. Indeed, there is no reason to expect that a frequently occurring photospheric condition is described by the same statistics as a rarely observed class of coronal events. To make the comparison of the photospheric and coronal events more consistent, one needs to make sure that the detected populations of events have approximately the same size -- in other words, they are described by {\it the same degree of intermittency}. 

Based on the population sizes provided in Table \ref{table1}, the $p=99.0 \%$ ($p=99.5\%$) magnetic events have approximately the same degree of intermittency as the $p=95.0\%$ ($p=97.0\%$) heating events. Pairwise comparison of these sets of events (99.0$\%$ (99.5$\%$) MDI threshold vs. 95.0$\%$ (97.0$\%$) EUVI threshold) shows that they are described by drastically different geometric exponents but surprisingly close distribution exponents, especially those of the time-integrated parameters $S$, $V$ and $E$ (see Table \ref{table2}). As we show in Section \ref{sec:coupling}, the distinct values of the linear scale distribution exponent $\tau_L$ and $\hat{\tau}_L$ in the photosphere and corona reflect substantially different geometries of the intermittent events in the two system.

Fig. \ref{fig7} presents a comparison of the probability distributions of MDI and EUVI events over the three measures $S$, $V$ and $E$ involving spatiotemporal integration. The distributions are built using the two pairs of percentile thresholds indicated above yielding consistent population sizes. To construct the distributions, we consider only the events whose linear sizes lie within the scaling interval of interest $(L_1, L_2)$. No transformations besides the standard probability density normalization were applied to the histograms. 

The MDI and EUVI distributions shown in Fig. \ref{fig7} are remarkably similar in shape and described by fairly close occurrence probabilities for the same parameter values. Within the statistical accuracy, they effectively coincide for both combinations of thresholds.

The similarity of coronal and photospheric distribution exponents is also evident from Fig. \ref{fig8} which shows time evolution of the exponents evaluated within a 90-minute moving window. The wide error bars reflecting the lack of the sufficient number of events prevent us from drawing solid conclusions regarding the observed trends, but it seems plausible that the probabilistic laws describing the two data sets are dynamically coupled. 

\subsection{Clustering of event locations}
\label{sec:CI}
Before proceeding with the analysis of scaling relations between the photospheric and coronal exponents, it is important to clarify the meaning of the linear size measure $L$ in the context of a more traditional correlation scale length describing solar plasmas.

In general, one can introduce two inherently different types of scale for describing spatially intermittent data sets like the ones represented by MDI or EUVI events. The first type of scale characterizes linear sizes of individual events. In our analysis, $L$ serves precisely this goal by measuring the spatial extent of each spatiotemporal event. The second type of scales is associated with the mutual arrangement of many such events into spatially correlated patterns. Such ensemble-based scaling behavior is not reflected by the individual event parameters defined by section \ref{sec:data_methods} and requires an additional investigation.

An efficient tool of analyzing spatial correlations between point objects is correlation integral statistics. The correlation integral (CI) is the second-order moment of the multifractal expansion of a complex correlated pattern into a set of generalized fractal dimensions \citep{grassberger83, grassberger83a}; see \citet{schuster05} for a pedagogical introduction to the concept. In a nutshell, the CI is the probability to find a pair of point objects within a hypersphere of a specified radius $r$, as a function of $r$. For a randomly scattered set of points, this probability scales as $\propto r^d$, where $d$ is the embedding (Eucledean) dimension of the set. For a clustered set, it goes as $\propto r^{\alpha_c}$ with the correlation dimension $\alpha_c < d$ up to some upper scale representing the correlation radius of the set and/or the finite size effects. The concept of CI can be generalized to study mutual correlations between two distinct data sets \citep{kantz94}, in which case the power-law index can be smaller or greater than $d$ depending on the sign of cross-correlations. In our preceding work  \citep{uritsky12}, we have successfully applied the CI method to investigate dynamic correlation properties of solar magnetic network imprinted in the locations of ``origin'' and ``demise'' magnetic events occurring when a compact group of pixels of a LOS magnetogram exceeds or goes below a detection threshold. 

A similar technique can be used to characterize spatial correlations of the MDI and EUVI events studied in the present paper.

Fig. \ref{fig9} shows CI statistics for two combinations of MDI and EUVI thresholds producing comparable amounts of events. To construct each CI plot, we considered average solar coordinates of the initial positions of the events. The CI counting statistics included all pairs of such origin points occurring simultaneously with respect to each other within the $\pm 300$ s ``grace'' period accounting for the granulation noise and $p-$mode contamination which limit temporal accuracy of photospheric observations \citep{deforest07, lamb08, uritsky12}. 

The power law slopes of all the CI plots presented in Fig. \ref{fig9} are statistically significantly below the value of the embedding dimension ($d=2$) corresponding to the null uncorrelated model. The data reveal the presence of correlations across the entire studied range of distance $r$ spanning from $\sim 3$ Mm up to $\sim 300$ Mm both for magnetic and heating events. This range of scales attributed to spatially correlated ensemble behavior of the events is much broader than the range of the linear sizes (3 to 12 Mm) of the events which were included in the analysis. 

The plots shown in Fig. \ref{fig9} suggest that despite their compact sizes, the events in our data sets carry information about a much larger process, most likely a superposition of the photospheric super- and meso-granulaiton.

\citet{simon95, simon01} developed an illustrative ``corks'' model of photopsheric supergranulation which can be invoked to clarify the discussed difference between the spatial scales $L$ and $r$. In this statistical model, the supergranulation network is generated by kinematic motion of elementary flux tubes (corks) transported passively by a supergranular flow modulated by random displacements mimicking granulation. In terms of this model, the scale $L$ reflects the finite linear size of individual flux tubes of the distance they travel during their lifetime $T$, whichever is greater. In a real-life photosphere, $L$ could be a footprint of a combination of secondary processes driven by the horizontal convection, such as e.g. magnetic cancellation, fragmentation and coalescence, superposed with the vertical emergence and submergense of solar magnetic bipoles. In contrast, the distance $r$ between the events is the correlation scale describing the mutual arrangement of many magnetic elements which is directly shaped by the flow. This scale is likely to reflect the global landscape of the turbulent photospheric flow rather than the time history of individual magnetic concentrations. 

Therefore, the scale measures $L$ and $r$ describe two aspects of the same driving process, the photospheric convection. 
This duality is important because it indicates that the seemingly narrow range of linear event sizes covered by our analysis is driven by a photospheric process involving a much broader range of scales.


\section{Scaling relations}
\label{sec:scaling_relations}


We now take a closer look at the spatiotemporal geometry of magnetic and emission events in order to derive scaling relations between the observed scaling exponents. 
The relations will be validated using the averaged scaling exponents describing MDI events detected at $p=99\%$ and $p=99.5\%$ thresholds, as compared to the averaged exponents of EUVI events obtained respectively at $p=95\%$ and $p=97\%$ (Table \ref{table3}). As mentioned in the previous section, these events are described by comparable levels of intermittency resulting in consistent population sizes. While the numerical values of the exponents vary with the detection threshold (Table \ref{table2}), the relations between the exponents  discussed in this section are expected to be insensitive to the threshold as long as the MDI and EUVI events are described by a similar degree of intermittency. 

As before, we use the hat symbol ($\hat{ \ }$) to distinguish coronal parameters from their photospheric counterparts. 

\subsection{Scaling of magnetic events}

According to the accepted scaling model (\ref{eq0b}), the scaling of linear size of the event with the event duration is governed by the geometric exponent $D_T$:
\begin{equation}\label{m1}
L \propto T^{1/D_T} = T^H,
\end{equation}
Since  $D_T \approx 2.18$ (Table \ref{table3}), the diffusive exponent $H \approx 0.46$ is close to the value 1/2. This means that the linear size of the perturbation scales approximately as a square root of the elapsed time. Such scaling is characteristic of classical (as opposed to fractional) random walk \citep{mandelbrot68} which can be a manifestation of random footpoint shifting of the photospheric field (see e.g. \citet{simon95,simon01,rast03,crouch07}). The idea of the random footpoint shuffling has been recently used for constructing a statistical fractal-diffusive self-organized criticality model (FD-SOC) \citep{aschwanden12, aschwanden12a}. The standard FD-SOC model \citep{aschwanden12a} with $H=1/2$ predicts the geometric and distribution exponents which are fairly close to the exponents of the MDI events reported in Table \ref{table3}.

Assuming that the average magnitude of the horizontal convective flow associated with the shuffling is independent of $L$ within the studied range of spatial scales, we get a simplified picture (Fig. \ref{fig10}(a)) in which the swept area of the event scales approximately as $S \propto \ell \ T$, where $\ell \sim A^{1/2}$ is a measure of the instantaneous linear scale of the event. Therefore we have $S \propto A^{1/2}T \propto L^{\frac{D_A}{2} + D_T}$ which, taking into account $S \propto L^{D_S}$, yields 
\begin{equation}\label{m3}
D_S = \frac{D_A}{2} + D_T.
\end{equation}
By plugging in the measured values of $D_A$ and $D_T$ we get $D_S \approx 2.88$. The direct evaluation of $D_S$ based on the $S(L)$ regression leads to $D_S = 2.78 \pm 0.11$ showing that the relation (\ref{m3}) holds with a reasonable accuracy. 

It is important to clarify that the equation (\ref{m3}) is derived assuming that the instantaneous shape of magnetic elements is purely 2-dimensional. While this assumption is certainly not justified for high- resolution magnetograms featuring irregular magnetic geometry \citep{cadavid94}, it seems to be acceptable for describing SOHO MDI structures covering up to 10 Mm$^2$ at a time (the largest $A$ consistent with the chosen $L$ range, see Fig. \ref{fig5}) since such structures are too small to display identifiable fractal boundaries at the available resolution.

The exponent $D_V$ describing the scaling of spatiotemporal volume $V$ of magnetic events can be also derived in frames of the random shuffling model implying that $V \sim A \ T$ (since the cross-section area $A$ is perpendicular to the time axis, see Fig. \ref{fig10}(a)). Making use of $A \propto L^{D_A}$ and $T \propto T^{D_T}$ we get 
\begin{equation}\label{m4}
D_V = D_A + D_T.
\end{equation}
The relation (\ref{m4}) predicts $D_V \approx 3.57$ which agrees well with the observed value $D_V = 3.43 \pm 0.12$.

The dependence of the instantaneous linear size $\ell$ of magnetic elements on the lifetime $T$ reflects the dynamic behavior of the underlying physical processes acting at small spatial scales, such as e.g. magnetic flux coalescence, fragmentation, emergence and submergence \citep{lamb08, chaouche11, crouch07, simon95, simon01, welsch06}. Even though these processes are not fully resolved in the studied data, their effective contribution can be evaluated based on the temporal scaling of magnetic element sizes as expressed by the measured geometric exponents. By combining $\ell \propto A^{1/2}$, $A \propto L^{D_A}$, and $L \propto T^{1/D_T}$ we obtain 
\begin{equation}\label{m5}
\ell \propto T^{H_1}, \ H_1 = \frac{D_A}{2 D_T}.
\end{equation}
Here, the diffusion index $H_1$ characterizes a small-scale transport of the photospheric magnetic field. Substituting $D_A \approx 1.39$ and $D_T \approx 2.18$ we get $H_1 \approx 0.32$. The obtained index shows that the small-scale magnetic field convection is strongly subdiffusive \citep{bouchaud90, isichenko92}, meaning that the size of the magnetic elements varies with time significantly slower than it would be expected for a regular diffusion described by the classical Brownian walk defined with $H_1 = 1/2$ (linear displacement proportional to the square root of the elapsed time). A subdiffusive transport usually takes place when the studied system contains spatial subdomains where the transported quantity becomes partially ``trapped'' \citep{metzler00}. In solar photosphere, such trapping can be due to a small-scale turbulence accompanying the granulation process \citep{espagnet93} which is closely related to the surface dynamo problem \citep{vogler07}. According to Kolmogorov's refined similarity hypothesis, the third moment of the distribution function of correlated velocity increments $\Delta u$ observed at the characteristic distance $\ell$ scales as $\ell ^{3 H_1}$ \citep{stolovitzky94}. Comparing this expression with Kolmogorov's exact relation for inertial range turbulence \citep{monin75, biskamp03} $\left\langle \Delta u (\ell) ^3 \right\rangle  = - \frac{4}{5} \left\langle  \epsilon \right\rangle  \ell$ ($\epsilon$ being the dissipation rate per unit mass) we arrive at $H_1 = 1/3$. This is fairly close to the $H_1$ value measured in the photosphere, showing that the range of the instantaneous sizes of the magnetic elements addressed by our analysis ($\ell \equiv A^{1/2} \sim  1.4 - 3$ Mm for 3Mm $\leq L \leq$ 12 Mm, see Fig. \ref{fig5} ) is likely to fall into the inertial range of the granulation-driven turbulence \citep{espagnet93}, and if so, the photospheric turbulence is locally isotropic.

\subsection{Scaling of emission events}

In contrast to the photospheric events evolving approximately as $L \propto T^{1/2}$, temporal dependence of coronal event sizes is strongly superdiffusive as revealed by $\hat{H} = 1/\hat{D}_T \approx 0.74$. The small-scale coronal diffusive exponent $\hat{H}_1 = \hat{D}_A / (2 \hat{D}_T ) \approx 0.59$ is also larger than the classical value. The superdiffusive behavior is known to arise in  strongly nonlinear media containing spatially distributed regions in which the diffusive transport is anomalously fast \citep{metzler00}. Compared to the classical random walk with temporally uncorrelated increments, the superdiffusive walk, such as the one modeled by fractional Brownian motion \citep{mandelbrot68}, features significant positive correlations across all time scales involved in the process \citep{turcotte89, feder88}. The fact that $\hat{H} \neq \hat{H}_1$ suggests that the dynamics of coronal emission regions is a sum of two types of motion -- a superdiffusive expansion of the event area and a displacement of the center of the event over a comparable distance (Fig. \ref{fig10}(b)). 

As the next step, we look further into the growth and decay dynamics of bright coronal regions characterized by geometric and distribution exponents. 

The expansive superdiffusive growth of emission regions could be a manifestation of energy avalanches developing in the corona (see e.g. \citet{aschwanden11} and references therein). Avalanching dynamics can be seen as a special case of spreading dynamics in active nonlinear media with spatially distributed energy sources and sinks \citep{vespignani98, munoz99}. Spreading experiments provide the most accurate determination of the critical point of systems with multiple absorbing states \citep{grassberger79} such as avalanching systems at self-organized criticality \citep{vespignani00}. A small perturbation representing an initial instability is created at the origin of an otherwise absorbing configuration, leading to a spread of activity. At the critical point associated with power-law distributions of avalanche sizes \citep{bak87, robinson94, sethna01}, the mean-squared deviation from the origin $R^2$, the survival probability $P$, and the number $N$ of unstable sites  scale with time $t$ as $R^2(t) \propto t^{\hat{z}}$, $P(t) \propto t^{- \hat{\delta}}$, and $N(t) \propto t^{\hat{\eta}}$ where $\hat{z}$, $\hat{\delta}$ and $\hat{\eta}$ are the spreading exponents describing the growth and decay of energy avalanches. 

In our measurements, $P$ has the meaning of the cumulative distribution of avalanche lifetimes: $P(t) = \int^{t}_{0}f(T)dT \propto T^{-\hat{\tau}_T + 1}$ and hence $\hat{\delta} = \hat{\tau}_T - 1$ \citep{uritsky07, morales08}. The number of individual unstable events (toppling events in sandpile simulations \citep{bak87}, nanoflares in the Parker's heating scenario \citep{charbonneau01}) associated with an avalanche with a lifetime $T$ scales as $ \propto T ^ {1 + \hat{\eta} + \hat{\delta} } $ \citep{munoz99, uritsky07, morales08}. Approximating this number by the integrated emission flux $E$ we obtain 
$E \propto T^{ 1 + \hat{\eta} + \hat{\delta} } = L^{ \hat{D}_T(1 + \hat{\eta} + \hat{\delta}) } \propto L^{\hat{D}_E}$ 
which yields 
\begin{equation}\label{c1}
1 + \hat{\eta} + \hat{\delta} = \frac{\hat{D}_E}{\hat{D}_T}.
\end{equation}

\citet{munoz99} have argued that the relations between the spreading and the probability distribution exponents depend upon the universality class of an avalanching system. 
Here we accept their interpretation in terms of the directed and dynamical percolation classes governed by the well-known relation $\hat{z}d/2 = \hat{\eta} + 2\hat{\delta}$ (see \citet{munoz99} and references therein) which in our case can be rewritten as
\begin{equation}\label{c2}
\hat{z}d/2 = \frac{\hat{D}_E}{\hat{D}_T} + \hat{\tau}_T - 2
\end{equation}
where $d$ is the Euclidean dimension of coronal avalanches. 
Also, from the definition of the $\hat{z}$ exponent it follows that  
\begin{equation}\label{c3}
\hat{z} = \frac{\hat{D}_S}{\hat{D}_T}.
\end{equation}
where we took into account that $R^2 \sim S$ and $t \sim T \propto L^{\hat{D}_T}$. By solving eqs. (\ref{c2}-\ref{c3}) for $d$ and substituting the measured coronal exponent values we find that $d \approx 3$, with the value $\hat{z} \approx 1.47$ predicted by eq.(\ref{c3}) being in an agreement with the behavior of three-dimensional dynamical percolation \citep{munoz99}. 

The derived three-dimensional geometry of energy avalanches in the corona is consistent with other scaling indices. The swept area $S$ of each event can be considered as a two-dimensional projection of a $d$-dimensional shape. The dimension $d_p$ of the projection  (approximated in our case by $\hat{D}_S$ by the definition of fractal dimension \citep[e.g.,][]{turcotte97}) is related to the dimension of the projected object as $d_p = d - 1$ \citep{mandelbrot82}. For $d = 3$, the relation predicts $\hat{D}_S = 2$ which matches exactly the measured value $2.00 \pm 0.03$ of this coronal exponent. Also, assuming (as before) that the integrated emission flux $E$ is proportional to the number of the unstable coronal sites constituting an avalanche, we expect $E \propto L^{\hat{D}_E}$ with $\hat{D}_E = 3$, which is in an agreement the empirical result $\hat{D}_E = 3.06 \pm 0.05$. 

In coronal heating studies, the value $d = 3$ is often taken for granted \citep{aschwanden00b, parnell00, krucker98,mitra_kraev01}. The analysis conducted here justifies this choice speaking in favor of a three-dimensional emitting plasma volume, as opposed to more complex geometries described by $d < 3$ as predicted by some statistical-physical models (see \citet{mcintosh01} and refs therein). Elementary instabilities associated with individual nanoflares can, in principle, have a different dimensionality, for instance $d = 2$ for reconnecting current sheets. However, our estimates show that the interaction network controlling growth and decay of coronal emission regions is likely to be embedded in a 3-dimensional space, and is also space-filling \citep{mandelbrot77}.

\subsection{Stochastic coupling relations}
\label{sec:coupling}

Our observations indicate that the MDI and EUVI events are described by rather similar probability distributions of $S$, $V$ and $E$ parameters but considerably different geometric scaling laws yielding distinct sets of $D_x$ and $\hat{D}_x$ exponents (see Table \ref{table3}). This subsection aims to resolve the apparent controversy by connecting these power law exponents through a set of scaling relations describing photosphere - corona interactions.  In section \ref{sec:energy}, these relations are used to provide a new insight into the energy budget of  multiscale photosphere - corona coupling. 

As the starting point, consider the subsets of coronal and photospheric events characterized by the same probability of occurrence. Since the probability distributions of the swept areas of coronal and photospheric events almost coincide, this requirement is equivalent to $S = \hat{S}$. 

Eq. (\ref{eq0b}) suggests that the linear scales describing the considered events are related as $L^{D_S} \propto \hat{L}^{ \hat{D}_S }$ yielding 
\begin{equation}\label{pc1}
\hat{L} \propto L^\alpha, \ \alpha = \frac{D_S}{\hat{D}_S} \ \approx 1.39,
\end{equation}
meaning that the apparent linear sizes of the coronal events are larger for the same swept area. The new scaling index $\alpha$ couples linear scales of equally probable coronal and photospheric events. The fact that $\alpha > 1$ means that the characteristic length of coronal emission regions grows faster with the swept area compared to the length scale of photospheric events. The difference in the scaling behavior of the two solar regions should hold under any linear transformation maintaining $S \propto \hat{S}$. 

Next, consider the lifetimes of the selected subsets of events described by $S = \hat{S}$. Since $T \propto L^{D_T}$ and $\hat{T} \propto \hat{L}^{\hat{D}_T}$, we have a similar coupling relation for the lifetimes of the coronal and photospheric events, with its own power-law index $\beta$:
\begin{equation}\label{pc2}
\hat{T} \propto T^\beta, \ \beta = \frac{D_S \hat{D}_T}{D_T \hat{D}_S} \ \approx 0.87.
\end{equation}
The value of coupling index $\beta$ indicates that the duration of the coronal emission events increases slower with the event area compared to the duration of the photospheric events. 

The relations (\ref{pc1})-(\ref{pc2}) suggest that the observed similarity of $f(S)$ and $f(\hat{S})$ distributions is achieved through a nontrivial rescaling of spatiotemporal sizes of solar events, with the coronal emission events lasting for a shorter time but propagating over a larger distance compared to the photospheric shuffling events described by the same occurrence rate. Our goal now is to derive coupling relations expressing coronal geometric and distribution exponents as functions of the photospheric exponents and the empirical indices $\alpha$ and $\beta$. 

By definition, the linear scale geometric exponent is identical to unity in both data sets, and so  
\begin{equation}\label{pc3_}
\hat{D}_L = D_L.
\end{equation}

The exponent $\hat{D}_S$ is already connected with $D_S$ by the scaling index $\alpha$ (\ref{pc1}). It can be also linked with other photospheric indices for checking their consistency. By combining the swept area expression for the photosphere (\ref{m3}) with the coupling relation (\ref{pc1}) we obtain $S \propto (\hat{L}^{-\alpha})^{\frac{D_A}{2}+D_T}$. After equating with $\hat{S} \propto \hat{L}^{\hat{D}_S}$, this equation provides an alternative form for the coronal exponent $\hat{D}_S$:
\begin{equation}\label{pc3}
\hat{D}_S = \frac{D_S}{\alpha}= \frac{D_A + 2 D_T}{2 \alpha}.
\end{equation}
The relation \eqref{pc3} predicts $\hat{D}_S \approx 2.0$ which coincides with the measured value for this exponent ($2.00 \pm 0.03$). 

Similarly, by plugging (\ref{pc1}) and (\ref{pc2}) into $\hat{T} \propto \hat{L}^{\hat{D}_T }$ and substituting $L^{D_T}$ for $T$ we obtain
\begin{equation}\label{pc4}
\hat{D}_T = \frac{\beta}{\alpha}D_T
\end{equation}
which yields the correct result $\hat{D}_T \approx 1.36$ as expected from the definition of $\alpha$ and $\beta$.

To derive the exponent $\hat{D}_A$, one should take into account a fractal (non-integer) dimension $d_F$ of the emission regions defined by $\hat{A} \propto \hat{\ell}^{d_F}$. In frames of our model (Fig. \ref{fig10}(b)) the area $\hat{S}$ covered by an emission event results from the combination of an avalanche-like expansion of the unstable region and a systematic displacement of its centroid. The characteristic linear scales of these effects are given respectively by $\hat{\ell}$ and $\hat{L}$ from which it follows that $\hat{S} \propto \ \hat{L} \hat{\ell} = \hat{L}^{1 + \frac{\hat{D}_A}{d_F}}$ and therefore 
\begin{equation}\label{pc5_}
d_F = \frac{\hat{D}_A}{\hat{D}_S - 1}. 
\end{equation}
The obtained relation allows us to evaluate the fractal dimension $d_F \approx 1.6$ without performing direct scaling analysis which would likely to be inaccurate due to the smallness of $\hat{\ell}$ and limited spatial resolution of EUVI images. Using (\ref{pc5_}) along with the coupling relation (\ref{pc3}) for $\hat{D}_S$  we get
\begin{equation}\label{pc5}
\hat{D}_A = \hat{d}_F \left( \frac{D_S}{\alpha} -1 \right) .
\end{equation}
The obtained expression matches the measured value $\hat{D}_A = 1.60 \pm 0.02$, which follows from the definition of $\alpha$ and $d_F$.

The exponent $\hat{D}_V$ can be related to $D_V$ by using the similarity of probability distributions of  spatiotemporal volumes of coronal and photospheric events (Fig. \ref{fig7}), by analogy with the analysis of the exponents  and $D_S$ presented above. For a given occurrence rate, $\hat{V} = V$ and hence $\hat{L}^{\hat{D}_V} \propto L^{D_V}$, which after applying the spatial rescaling (\ref{pc1}) leads to
\begin{equation}\label{pc6}
\hat{D}_V = \frac{D_V}{\alpha}.
\end{equation}
Comparing (\ref{pc3}) and (\ref{pc6}) one can easily notice that $ ( \hat{D}_S D_V ) / (\hat{D}_V D_S)  = 1$. This theoretical identity is approximately fulfilled: the substitution of the measured exponents yields $\sim$0.95. 

Expressing the coronal exponent $\hat{D}_E$ via the photospheric exponents is a critically important (albeit somewhat more challenging) step because $\hat{D}_E$, along with the distribution exponent $\hat{\tau}_E$, can be used for forecasting the level of EUV coronal emission for a given photospheric condition. Using the relations (\ref{c2}) and (\ref{c3}) for the coronal spreading exponent $\hat{z}$ along with the coupling relations (\ref{pc3}) and (\ref{pc4}), after some algebra we arrive at 
\begin{equation}\label{pc7}
\hat{D}_E = \frac{\beta D_T}{\alpha} \left( \frac{d \ D_S }{2 \beta D_T} - \hat{\tau}_T \right).
\end{equation}
Note that this is an intermediate result because it depends on the coronal exponent $\hat{\tau}_T$ which is yet to be derived. 

The general form for any distribution exponent $\hat{\tau}_x$ follows directly from the probability conservation condition $f(\hat{X})d\hat{X} = f(X)dX$ where $\hat{X}$ and $X$ stand for the same measures in the corona and photosphere. Assuming a power-law scaling $\hat{X} = X^{\gamma_x}$ between the two, and applying the power-law model (\ref{eq0a}), we get $f(\hat{X}) d\hat{X} \propto X^{-\gamma_x \hat{\tau}_x} X^{\gamma_x - 1} dX \propto X^{-\tau_x} dX $ and consequently
\begin{equation}\label{pc8}
\hat{\tau}_x = \frac{\tau_x - 1}{\gamma_x} +1, \ \gamma_x = \frac{\hat{D}_x}{D_x}\alpha,
\end{equation}
in which the expression for $\gamma_x$ follows from $\hat{X} \propto \hat{L}^{\hat{D}_x} \sim (L^{\alpha})^{\hat{D}_x} \sim ((X^{1/D_x})^{\alpha})^{\hat{D}_x} \sim X^{ \frac{\hat{D}_x} {D_x} \alpha }$. By using the coupling relations for the geometric exponents derived above, one can easily calculate the corresponding value of $\gamma_x$ and on this basis $\hat{\tau}_x$. In particular, for the linear size distribution exponent we have $\gamma_L = \alpha$ (since $\hat{D}_L = D_L \equiv 1$) and 
\begin{equation}\label{pc9}
\hat{\tau}_L = \frac{\tau_L - 1}{\alpha} +1.
\end{equation}
In its turn, the lifetime distribution exponent can be derived in a similar fashion by taking into consideration eq.(\ref{pc4}) which yields $\gamma_T = \beta$ and therefore
\begin{equation}\label{pc10}
\hat{\tau}_T = \frac{\tau_T - 1}{\beta} +1.
\end{equation}
The relations (\ref{pc9}) and (\ref{pc10}) result in the estimates $\hat{\tau}_L \approx 2.55$ and $\hat{\tau}_T \approx 1.91$ which agree well with the measured values $\hat{\tau}_L = 2.57 \pm 0.13$ and $\hat{\tau}_T = 2.02 \pm 0.07$.

Finally, after the substitution of (\ref{pc10}) into (\ref{pc7}) we obtain the geometric exponent of the emission flux as a function of the photospheric scaling exponents and the empirical coupling indices $\alpha$ and $\beta$:
\begin{equation}\label{pc11}
\hat{D}_E = \frac{d D_S /2 + (\beta - \tau_T +1)D_T}{\alpha}.
\end{equation}
For space-filling 3-dimensional emission regions ($d=3$) the relation (\ref{pc11}) predicts $\hat{D}_E \approx 3.12$ which is approximately consistent with the observed $\hat{D}_E = 3.06 \pm 0.03$.

The last column in Table \ref{table3} shows the values of the coronal distribution exponents calculated based on the scaling relations discussed in this section.

\section{Discussion}
\label{sec:mechanisms}

Our findings indicate that multiscale energy dissipation in the corona is closely coupled with the dynamics of the underlying photosphere. The derived scaling laws suggest a coupling scenario in which the footpoints of the coronal flux tubes undergo quasi one-dimensional horizontal convective motions driven by a vector sum of uncorrelated displacements (a random walk behavior) while the coronal emission events are three-dimensional, with the growth and decay dynamics being similar to that of the avalanching models. The physical mechanism mediating the interaction between the two solar regions remains to be clarified.

\subsection{Routes to multiscale dissipation}

Multiscale stochastic coronal dissipation is usually considered in frames of two statistical-physical frameworks, SOC and turbulence. Both approaches deal with open dissipative systems containing energy sources and sinks and can incorporate a structured photospheric driver.

The first of the two frameworks, self-organized criticality \citep{bak87}, has been extensively explored in the context of solar flares starting from the pioneering work of \citet{lu91}. SOC seeks to explain power-law distributions of flare parameters via cooperative interactions of a large number of nonlinearly coupled degrees of freedom representing unstable coronal loops and/or loop strands \citep{crosby93, crosby11, charbonneau01, aschwanden00b, aschwanden02, morales08, uritsky07}. It has been suggested that the nonpotential magnetic field configurations existing in the corona release their free energy through a chain interaction of multiple spatially localized instabilities such as those associated with nanoflare heating \citep{parker88, klimchuk06, viall11}. The perturbation proceeds explosively by involving a growing number of unstable plasma regions, similarly to the interaction of rolling grains of sand leading to a sand pile avalanche \citep{bak87, bak88}. 

Coronal energy avalanches are not always topologically compact. The super-diffusivity of solar corona can be a manifestation of multiple spatially separated dissipation regions appearing in response to the same initial trigger \citep{torok11}. The ensemble-based behavior of these regions would involve spatial jumps (such as those mimicked by L\'{e}vy flights) associated with the appearance of new centers of activity, giving rise to a superdiffusive $\hat{H}$ exponent. The evolution of individual regions could follow a different scaling law and be subdiffusive during the impulsive phase of flares \citep{aschwanden12}.

One of the necessary theoretical conditions for the SOC state is time scale separation between the driver, the avalanches, and the individual instabilities constituting the avalanches \citep{vespignani98}. Most of the numerical SOC models operating in this mode are not solvable analytically \citep{jensen98}, making it difficult to obtain closed-form solutions for observable scaling laws. The first-principle analysis of these models is complicated by the involvement of higher-order statistical moments \citep{vespignani98, sethna01} requiring full dynamic renormalization group treatment \citep{chang92}. If the behavior of solar corona was fully analogous to such models, the free energy lanscape in the corona would be solely formed by preceding flaring events independently of the driver, as exemplified by non-Abelian sandpile models \citep{hughes02, uritsky04}. 

However, coronal dynamics considered in our paper is fundamentally different from such pure SOC behavior. Applying the above time scale separation principle to the Sun, the correlation time of the photospheric input needs to be significantly larger than the lifetime of the flaring events which, in their turn, should be much longer than the relaxation time of the nanoflares. While the latter condition is likely to be met, the former one is definitely not. In fact, our study deals with the photospheric events whose relaxation times ($T = 8 \times 10^2 - 1.2 \times 10^4 $ s) overlap with the coronal time scales ($\hat{T} = 3 \times 10^2 - 2 \times 10^3 $ s), see Fig. \ref{fig5}. In such regime, correlation properties of the driver are rather essential for the resulting stochastic dynamics. Since the spatial distribution of free magnetic energy supplied by the photosphere is far from random \citep{uritsky12}, the resulting heating should be a \textit{driven} rather than a fully self-organized process. Our study supports this interpretation by showing that solar photosphere places clear-cut constraints on the occurrence probability of coronal events of various sizes, even though their immediate triggers can be unpredictable \citep{aschwanden11, aschwanden13}. Another indication of a driven regime is that numerical values of coronal critical exponents do not fall into any known universality class of SOC dynamics. Some of the exponents such as $\hat{\tau}_T$ and $\hat{\tau}_E$ are consistent with previously studied paradigmatic avalanching systems \citep{corral99, baiesi06, charbonneau01}, but others are not. In particular, we are not aware of any sandpile symmetry group that would produce the combination $\hat{D}_S \sim 2.0$ and $\hat{D}_E \sim 3.0$ implying a 2-dimensional avalanche shape and a 3-dimensional dissipation.

The second framework commonly used for describing mutiscale coronal complexity is the high-Reynolds number magnetohydrodynamic turbulence, see e.g. \citet{biskamp03} and references therein. According to this scenario, multiscale dissiaption events in the corona result from turbulent intermittency (inhomogeneous energy dissipation) associated with a cross-scale energy cascade \citep{boffetta99, nigro04}. The energy is initially injected at the largest scale is due to e.g. photospheric motions. Next, nonlinear effects in the corona transfer this energy to ever smaller scales, similarly to the generation of multiscale eddies in turbulent fluids \citep{dmitruk97, buchlin03, nigro04, buchlin03, buchlin05, buchlin07, rappazzo10}. Intermittent turbulence can generate SOC-like probability distributions of dissipative events coexisting with traditional fluid signatures such as power-law scaling of structure functions \citep{uritsky07}. 

\citet{dmitruk97} used an externally driven MHD system to study the dynamics of transverse sections of coronal loops. Under a large-scale magnetic forcing, their system reached a turbulent state characterized by am algebraic distributions of the dissipated energy with a power law slope of $\sim 1.5$. This slope coincides, within statistical uncertainty, with our estimate of the exponent $\hat{\tau}_E$ which varies between 1.47 and 1.58 depending on the threshold (Table \ref{table2}).  The nanoflare model presented by \citet{nigro04} was based on the shell technique in the wave vector space applied to the set of reduced MHD equations. Their simulations showed that the energy injected due to photospheric footpoint motion is efficiently stored in the coronal loop giving rise to significant magnetic and velocity fluctuations whose interaction results in a turbulent cascade. This shell operates in a strongly intermittent regime which is qualitatively similar to the one studied by \citet{dmitruk97}, with dissipation peaks separated in time by quiet periods. The energy distribution power-law exponent $\sim 1.8$ obtained by \citet{nigro04} is somewhat higher than our estimates, while the lifetime exponent ($\sim 2$) is significantly larger than the observed $\hat{\tau}_T = 1.28 - 1.38$. 

\citet{buchlin05} and \citet{buchlin07} investigated probability distributions of dissipative events generated by MHD turbulence. \citet{buchlin05} conducted a set of numerical tests which showed that the power-law slopes of the event distributions are sensitive to the definition of the low-energy events generated by a shell-model MHD turbulence. The statistics of the high-energy events similar to the ones studied in our work were found to be more robust. In essence, our event definition is a 2D version of the ``threshold-background'' detection method described by \citet{buchlin05}. However, the lifetime distribution exponent that we obtained is significantly lower than the value observed in their simulation ($\sim 2.5$). The power-law index of energy distribution is close to our estimates, but the range of scales of the simulated power-law behavior is quite narrow suggesting a log-normal probability decay characteristic of toy turbulence models \citep{uritsky09}. A more advanced MHD model of coronal loop turbulence was developed by \citet{buchlin07}. The model consists of multiple ``shells'' representing different loop cross-sections coupled together through a strongly stratified longitudinal B-field. The lifetime exponent value $\approx 1.7$ obtained in this geometry is larger than our $\hat{\tau}_T$ estimates while the power-law energy distribution slope ($\sim 1.7$) matches our results and extents over a broad range of scales.

The hierarchy of scales characterizing MHD turbulence in coronal loops can be generated self-consistently at the coronal altitudes independently of the properties of the photospheric driving. 
\citet{rappazzo10} performed reduced MHD simulations to shaw that coronal dynamics is largely insensitive to spatial correlations in the photosphere once the coronal plasma enters a fully developed turbulent state. They concluded that the stochastic behavior of coronal loops observed within the inertial range of scales is an intrinsic property of the MHD turbulent cascade and not that of the driver. \citet{buchlin03} arrived at a similar conclusion based on a simplified RMHD model with an on/off current driven instability and a turbulent photospheric driver. The shape of the Fourier spectrum describing the driver had no noticeable effect on the coronal energy distribution exponent which stayed in the range $1.6 - 1.8$ in all their runs. On the other hand, simulation results reported by \citet{van11} show that turbulent heating rate increases nonlinearly (almost quadratically) with the vorticity of the footpoint motion. If the vorticity is a function of spatial scale, which is typically the case in 2D and 3D hydrodynamic and MHD turbulence \citep{lundgren82, gilbert93, politano95, jimenez96, bouchet09, servidio09, uritsky10}, this dependence provides a natural scale-dependent driving mechanism across the inertial range of turbulent scales.

While the inertial regime of coronal turbulence may or may not be perceptive to the structure of the photospheric forcing, the sub-inertial behavior could well be directly driven. Our statistics speak in favor of such directly driven mode of the photosphere-corona coupling at the transverse coronal wave scales above $\sim$ 3-5 Mm which seem to be strongly affected by multiscale photospheric dynamics. 


A number of hybrid stochastic models combining properties of continuum turbulence and discrete SOC dynamics have been also proposed, see e.g. \citet{lu95, klimas00, uritsky01a, buchlin03, valdivia03, klimas04}. The results of these simulations suggest that in order for a high-Reynolds number fluid to reach the state of SOC, the behavior of the fluid should involve two distinct modes of energy transport with well-separated time scales (by analogy with stick-slip transitions in SOC models), such as those associated with convective and diffusive magnetic energy transports acting at different scales. The main theoretical works in this potentially fruitful direction of the investigation of coronal complexity still lies ahead.

\subsection{Photospheric structuring}


When considering the origin of the photospheric structuring addressed by our study one needs to keep in mind that the linear scale $L$ represents the characteristic size of the magnetic concentration regions and is not a measure of the distance at which the regions are arranged with respect to each other (see Section \ref{sec:CI}). The inter-event distance spans over a much broader range of scales (see Fig. \ref{fig9}) encompassing different regimes of flow motion in the photosphere such as granulation, mesogranulaiton and supergranulation. 

The supergranular flow is known to be dominated by the horizontal component (e.g., Leighton et al. (1962); Hathaway et al. (2000)). It typically leads to the formation of the magnetic network outlining supergranular cell boundaries (Hagenaar et al. 1997; Schrijver et al. 1997; Srikanth et al. 2000). In deterministic models, the photospheric network is produced by a forced rearrangement of small-scale surface magnetic fields driven by this large-scale horizontal flow (Martin 1988; Lisle et al. 2000; Wang et al. 1996). In stochastic models, the supergranulation network is generated by a kinematic-like motion of elementary flux tubes convected passively by partly randomized supergranular flow (Simon et al. 1995, 2001), by mutual advection of thermal downflows occurring at random locations (Rast 2003), or as a self-organization of nonlinearly interacting magnetic elements  following a random walk pattern (Crouch et al. 2007). In any of the above scenarios, the scale of the flow can be substantially larger that the characteristic distance between the magnetic elements driven by the flow into the narrow lanes subdividing supergranular cells. 

The origin of photospheric mesogranulation, as well as  its ability to provide free energy for the corona, are more controversial. It has been suggested that mesogranular flows could correspond to a distinct physical process. \citet{lawrence01} have shown, based on the analysis of SOHO MDI Doppler images, that solar mesogranules can be associated with a particular type of convective cells with size scale $\sim 4$Mm and coherence time greater than 40 min, coexisting with a turbulent stochastic flow carrying about one third of the flow energy. The horizontal flows were found to exhibit temporal decorrelation over a wide range of spatial scales (4 to 120 Mm). Using flow divergence maps constructed from similar data, \citet{shine00} were able to detect an outward convection (advection) inside mesogranules within each supergranule surrounded by thin boundaries featuring negative divergence. 

More recent studies have challenged this interpretation by showing that mesogranulation can be an artifact of commonly used averaging procedures. For instance, \citet{matloch09} have argued that mesogranulation pattern at the solar surface can be solely due to statistical properties of granules and intergranular lanes. The apparent structure possesses no intrinsic length or time scales which would indicate the presence of a convection cell corresponding to mesogranules. A significant stochastic component of the horizontal photopsheric flow with similar statistical properties (no characteristic scale at the mesogranular level) has been reported by \citet{rieutord08} based on a granule tracking of CALAS high-resolution solar images. The physics behind these statistical effects can be nontrivial as they involve nonlinear interaction photosphic granules leading to their self-arrangement. 



\subsection{Energy considerations}
\label{sec:energy}

The statistics shown in Fig. \ref{fig7} are consistent with a picture in which the coronal emission events are generated in a random subset of unstable magnetic volumes created by the photospheric motion. This interpretation is in line with the analysis of unstable magnetic discontinuities conducted by \citep{vlahos04}. Using a linear force-free extrapolation of the photospheric field, they  identified a large number of magnetic discontinuities whose free magnetic energies and volumes obey power-law distribution functions. The log-log indices of the distributions were found to be in agreement with the indices describing the occurrence frequencies of solar flare energies reported in the literature. Based on this comparison, \citet{vlahos04} speculated that the statistics of flares may result from a preexisting free energy fragmentation in the solar corona. In frames of this model, it is not expected that every unstable magnetic volume generates a flaring event, but most of the flares do occur in such pre-existing volumes containing strong nonpotential magnetic field. As we show below, our results are consistent with this picture but they do not rule out a possibility of an AC coupling mechanism involving multiscale pulses of upwardly propagating Alfv\'{e}nic waves \citep{suzuki06, ulrich96, fujimura09}. 


The normal Poynting flux $S_n$ of energy through the base of the corona in the ideal magnetohydrodynamic approximation can be written as 
\begin{equation}\label{d1}
S_n = \frac{c}{4\pi}\ \mathbf{n} \cdot \left( - \frac{ \mathbf{v} \times \mathbf{B} }{c} \times \mathbf{B} \right)
\end{equation}
in which  $\mathbf{v}$ is the photospheric velocity and $\mathbf{n}$ is the unit normal vector (for simplicity the magnetic diffusivity is assumed to be negligible). A similar expression describes the potential magnetic energy flux; the difference between the two fluxes yields the flux of \textit{free} magnetic energy into the corona \citep{welsch06}: 
\begin{equation}\label{d2}
S_n^{(F)} = \frac{1}{4\pi} \left( \mathbf{B}_h - \mathbf{B}_h^{(P)} \right) \cdot \left( v_n \mathbf{B}_h - \mathbf{v}_h B_n \right),
\end{equation}
where the subscript $h$ ($n$) denotes the horizontal (normal) components of the velocity and magnetic field vectors and $\mathbf{B}_h^{(P)}$ is the potential horizontal field. The studied FoV is close to the center of the solar disc where SOHO MDI flux counts provide a reasonable proxy to $\mathbf{B}_n$. The behavior of the horizontal velocity can be evaluated indirectly based on the tracks of the magnetic features as discussed in section \ref{sec:scaling_relations}. 

Assuming that the difference between the total and the potential horizontal magnetic fields remains constant during each event and ignoring the contribution from the $v_n \mathbf{B}_h$ term which cannot be evaluated from the studied data, we find that on average,
\begin{equation}\label{d3}
S_n^{(F)} \propto \left\langle v_h \right\rangle \left\langle B_n \right\rangle
\end{equation}
with $\left\langle v_h \right\rangle$ and $\left\langle B_n \right\rangle$ being the characteristic values of the horizontal flow velocity and the line of sight magnetic field, respectively. 
To get \eqref{d3}, we made and assumption $ \left\langle v_h B_n \right\rangle \propto \left\langle v_h \right\rangle \left\langle B_n \right\rangle$ therefore neglecting any correlations between the horizontal motion and the vertical magnetic field during the lifetime of the event. This assumption can be justified if the B-filed behaves as a passive scalar \citep{warhaft00} and has no effect on the velocity field, while the volumes of the advected frozen-in magnetic flux tubes do not depend on local flow velocity. 
The sign of $S_n^{(F)}$ in eq. (\ref{d3}) is positive since we are not interested in the events producing a downwardly directed Poynting flux $S_n^{(F)}<0$. 

As we stated earlier, the dynamics of the photospheric footpoint shuffling is close to classical random walk with $H=1/2$. Shuffling dynamics has been shown to be critically important for the generation of the upward Poynting flux, especially in strongly magnetized intergranular lanes experiencing horizontal vortex motions \citep{shelyag12}. The velocity fields satisfying classical Brownian motion condition are described by  stable probability distributions with finite and constant first and second moments \citep{turcotte97}. Based on this model, one can assume that $\left\langle v_h \right\rangle$ remains constant during each event and is not a function of $L$ (such a dependence implies a super- or sub-diffusive behavior of magnetic events which is not observed). The quantity $\left\langle B_n \right\rangle$ can be easily obtained based on the conducted scaling analysis. The spatiotemporal sum of magnetic flux counts during the events scales as $\propto L^{D_E}$ while the spatiotemporal volume scales as $\propto L^{D_V}$. The ratio between these two quantities gives the required average field: 
\begin{equation}\label{d4}
\left\langle B_n \right\rangle \propto L^{D_E - D_V}
\end{equation}

The free magnetic energy $W$ produced by any given shuffling event can be evaluated by integrating eq. (\ref{d3}) over the involved surface area and the event lifetime:
\begin{equation}\label{d5}
W \propto \left\langle v_h \right\rangle \left\langle B_n \right\rangle \int\limits_{T} \! dt \!\!\!\!\! \int\limits_{x,y \in a(t)} \!\!\!\!\!\!\!\! dx dy  
\end{equation}
in which $a(t)$ is the instantaneous area of the event at time $t$. It is easy to verify that the domain of the integration in eq. (\ref{d5}) is nothing but the swept area $S$ of the event, according to our definition of this parameter. By making use of this fact and substituting (\ref{d4}) for the magnetic field we arrive at the geometric scaling law for the energy of a shuffling event:
\begin{equation}\label{d6}
W \propto L^{D_E - D_V + D_S}.
\end{equation}
Substitution of the estimated values of the photospheric scaling exponents involved in this relation yields $W \propto L^{D_W}$ with $D_W \approx 3.15$. This estimate is in a remarkably good agreement with the geometric exponent $\hat{D}_E  = 3.04 \pm 0.03$ characterizing the power-law scaling of the coronal emission flux with $L$. 
Therefore, the shuffling energy and the coronal emission energy in the populations of coronal and photospheric events described by the same rate of occurrence are \textit{directly proportional}. While the proportionality itself does not prove a causal link, it clearly speaks in favor of a photospheric input working as a driver of coronal dissipation. 

The discussed coupling scenario is reminiscent of the famous Parker's scenario of nanoflare coronal heating. However, the similarity can be misleading for the following reasons. First, the physical scales of the coronal events included in our analysis is incomparably larger than the size of the nanoflares. The detection threshold used to extract the events was tuned to a fairly high emission level which ensured that most of the low intensity coronal brightenings were excluded from the statistics. Secondly, if we postulate that the smaller events (unresolved by our technique) obey the same statistical laws as the observed events, the obtained value $\hat{\tau}_E \approx 1.5 < 2$ of the emission exponent predicts that the energy outputs from the small events (possibly including the nanoflares) is negligible compared to the contribution of large events since the ensemble averaged emission flux
\begin{equation}\label{d7}
\left\langle E \right\rangle \! = \!\!\! \int\limits_{E_{min}}^{E_{max}} \!\!\!\! E f(E) dE \propto \left.\ E^{2 - \hat{\tau}_E } \right|_{E_{min}}^{E_{max}} \sim (E_{max})^{0.5}
\end{equation}
is controlled by the largest flares. 
On the other hand, the ratio between the free magnetic energy of a given shuffling event (\ref{d6}) to the average total magnetic energy per unit area during the event increases with decreasing $L$:
\begin{eqnarray}\label{d8}
\frac{\mbox{Free magnetic energy}}{\mbox{Total magnetic energy}} \propto \frac{W}{\int_{V}B_n^2 \ dv } \\
\propto \frac{L^{D_E - D_V + D_S}}{L^{2(D_E - D_V) + 3}} = \frac{L^{3.15}}{L^{3.74} } = L^{-0.84}
\end{eqnarray}
in which we used the energy scaling relation (\ref{d6}) for $W$, the equation (\ref{d4}) for the average LOS magnetic field during the event, and have assumed a 3-dimensional coronal volume. The fact that the resulting power law exponent is negative indicates that the small-scale magnetic events are on average more effective as free energy sources compared to the large events. The higher relative occurrence rate of the small shuffling events could make this dependence statistically important. However, the net free energy outcome from the small photospheric events is nevertheless lower than that from the large events because the power-law exponent $\tau_W$ of the $W$ distribution predicted by the probability conservation ($D_W(\tau_W - 1) = D_L(\tau_L - 1) \Rightarrow \tau_W \approx 1.7 $) is, again, below the value 2. 

An alternative line of interpretation of the energy exchange between solar photosphere and corona involves multiscale Alfv\'{e}nic wave packets excited in the chromosphere and corona by the photospheric footpoint motions \citep[e.g.,][]{ofman98, ulrich96, kudoh99, suzuki06, depontieu07, fujimura09, ofman10}. Perturbations in the photosphere are able to generate a broad spectrum of transverse waves with the statistical properties of the generating convective events. Once these waves arrive at the corona they could result in dissipation events with a similar probabilistic pattern. 

In the Wentzel-Kramers-Brillouin approximation, the energy flux of an Alfv\'{e}n wave packet generated by a single shuffling event associated with the displacement velocity $v_d$ is given by 
\begin{equation}\label{d9}
S_n^{(A)} = \rho v_d^2 V_A
\end{equation}
in which $\rho$ is the ambient mass density and $V_A$ is the local Alfv\'{e}n speed. 

Assuming that the displacement velocity $v_d \approx \left\langle v_h\right\rangle$ is roughly constant (see the Brownian walk argument above) and the plasma is homogeneous, integration of (\ref{d9}) over the emitting area and time yields the Alfv\'{e}n energy
\begin{equation}\label{d10}
W_{A} \propto \left\langle v_h \right\rangle^2 \left\langle B_n \right\rangle \int\limits_{T} dt \!\!\!\!\!\! \int\limits_{x,y \in a(t)} \!\!\!\!\!\!\!\!\ dx dy 
\end{equation}
Since $\left\langle B_n \right\rangle \sim E/V$ and the domain of the integration is defined by the swept area $S$, we arrive at the scaling law
\begin{equation}\label{d11}
W_A \propto L^{D_E - D_V + D_S} \equiv L^{D_W}
\end{equation}
which is identical to (\ref{d6}). Thus, the AC heating mechanism leads to the same geometric scaling of the energy input from the photosphere as the DC mechanism based on the nonpotential MHD Poynting flux. Both mechanisms predict $D_W \approx \hat{D}_E$ and are in an agreement with the statistics of coronal energy dissipation presented in section \ref{sec:results}. 

It should be emphasized that the WKB approximation is generally valid for high frequency Alfv\'{e}n waves only. Non-WKB waves such as those studied by \citet{suzuki06} may exhibit a significantly different behavior leading to a more complex relationship between the generating convective events and the photospheric dissipation. 
The wave mechanism reflected by eq. (\ref{d11}) is more likely to play a role in the generation of short-living coronal events while the long-lasting events could be driven by a slow convective shuffling and/or non-WKB low-frequency waves and turbulence. Understanding the detailed interplay between these effects in the context of multiscale photosphere - corona coupling is an important task for future research.

\subsection{Back reaction on the photosphere}


Throughout this discussion, we have silently assumed that the photosphere drives the corona but is not influenced by its response. However, our analysis provides no constraints on the causal connection between the two solar region.

Due to the high gas pressure to magnetic pressure ratio in the photosphere, the latter is often considered as a one-way driver experiencing no coronal feedback. Since coronal field lines are attached to a dense photosphere, no rapid changes in the photospheric field associated with coronal eruptions are expected. Recent observations challenge this simplified picture showing that the photosphere - corona coupling can be a two-way process. The back reaction on the photosphere has been successfully identified after X-class flares \citep{wang10} and, more recently, for much smaller C-class flares \citep{wang12}.These findings are in line with the prediction by \citet{hudson08} that the photospheric magnetic fields should become more horizontal after flares despite the high plasma beta at the solar surface. Numerical simulation of the turbulent coronal loop heating have shown that the shape of the perpendicular wavenumber spectra of the corona can be imprinted in the photospheric spectra \citep{ van11, verdini12a} suggesting a possibility of a back reaction which is topologically much more complex. The back reaction on the photosphere is likely to be insignificant under  quiet solar conditions, but it could, to some extent, contribute to our statistical results.


\section{Conclusive remarks}
\label{sec:conclusions}

We presented a new statistical-physical approach to testing the photosphere-corona coupling as manifested in a quiet solar region. We investigated large ensembles of photospheric and coronal events detected in co-aligned sets of images provided respectively by SOHO MDI and STEREO EUVI instruments. 
The results obtained reveal a statistical consistency in the behavior of the two groups of events. For properly adjusted detection thresholds corresponding to the same degree of intermittency in the photosphere and corona, the events in the two solar regions exhibit nearly identical probability distributions of time-integrated measures (swept area, spatiotemporal volume, and integrated intensity). 

These findings suggest that multiscale photospheric motions shape the statistical distribution of solar flares, at least for a quiet Sun and for the range of spatial and temporal scales covered by our study (event sizes $L = 3-12$ Mm, event duraitons $T = 3 \times 10^2 - 2 \times 10^3 $ s). For this population of heating events, coronal activity is preconditioned by photospheric motions. As we have shown, \textit{the free energy input from the photosphere and the radiative energy output in the corona are fragmented into multiscale packets with matching power-law probability distributions}. 

The analysis of scaling relations between the obtained scaling indices has shown that the probabilistic behavior of the MDI and EUVI events is consistent with their characteristic spatiotemporal geometries and also indicates a strong physical interaction between the photosphere and corona. 

The derived scaling laws are in an agreement with the classical model in which the photospheric flux tubes undergo quasi one-dimensional horizontal convective motions driven by uncorrelated, random walk - like displacements. These shuffling displacements energize the corona across a broad range of scales and result in the power-law distributions of solar flares. 

The underlying physical mechanism of the photosphere-corona interaction cannot be rigorously validated based on our analysis alone. Most likely, it involves a variety of nonlinearly processes such as SOC, MHD turbulence, and wave-particle interactions. Our analysis shows that the dynamics of solar photosphere is an ultimate driver for any such process. It also provides a set of clear-cut observational constraints which can be taken into account in future studies. Our results suggest that each successful shuffling event generating a positive line of sight Poynting flux sends a packet of free energy into the corona via a multitude of DC and/or AC coupling mechanisms (see e.g. \citet{davila87, ofman94, ofman95, klimchuk06, mcintosh11}). The amount of the injected free energy associated with each shuffling follows a power-law distribution described by the same log-log slope ($\sim 1.5$) as the distribution of the integrated emission flux in the corona. The input and output energies also share the same geometric scaling exponent $\sim 3$.

It is worth mentioning that the wave coupling scenario can reach far beyond the corona and affect the acceleration and heating of the solar wind \citep[see e.g.][]{ofman10}.
The influence of a structured photosphere on spatial scaling properties of the solar wind is likely to be strongly nonlinear and in many cases dominated by locally generated turbulent structures and flows \citep{verdini12}. The frequency break separating the {\it in situ } portion of the transverse energy spectrum from the driven portion tends to shift toward lower frequency with heliocentric distance due to the solar wind expansion \citep{matthaeus86}.  The photospheric imprint can be completely lost when the spectrum is dominated by shocks \citep{suzuki06}, although rotational discontinuities which are found in many heliospheric current sheets can be controlled by Alfv\'{e}n waves propagating upward from the coronal base in a structured solar atmosphere, with DC motions being mainly responsible for the energy input \citep{malara12}. 

\begin{figure*}[htbp]
\includegraphics*[width=8 cm]{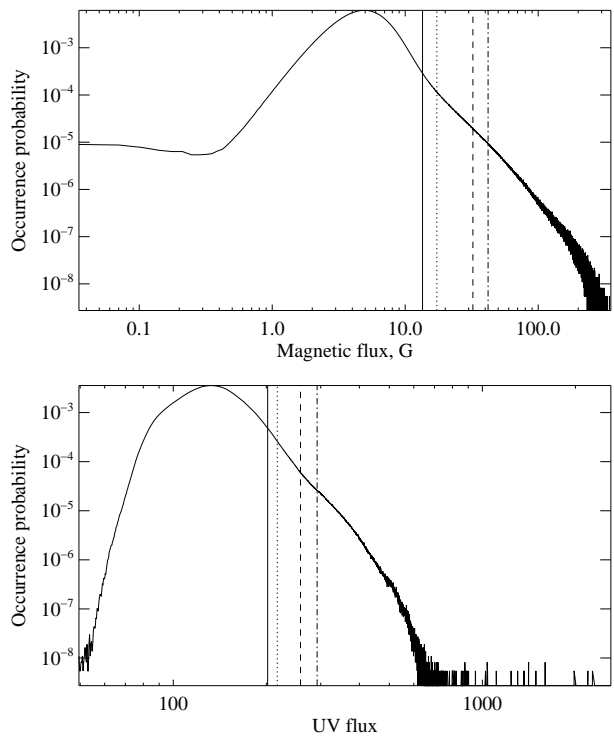} 
\caption{\label{fig3} Probability distributions of MDI magnetic fluxes (top) and EUVI emission fluxes (bottom) describing the entire interval of analysis (17:29:00 04/05/2007 - 10:58 04/06/2007). Solid, dotted, dashed and dot-dashed vertical lines show respectively the percentile levels $p = $95.0, 97.0, 99.0, and 99.5 $\%$ in each data set.}
\end{figure*}

\begin{figure*}[htbp]
\begin{center}
\includegraphics*[width=12 cm]{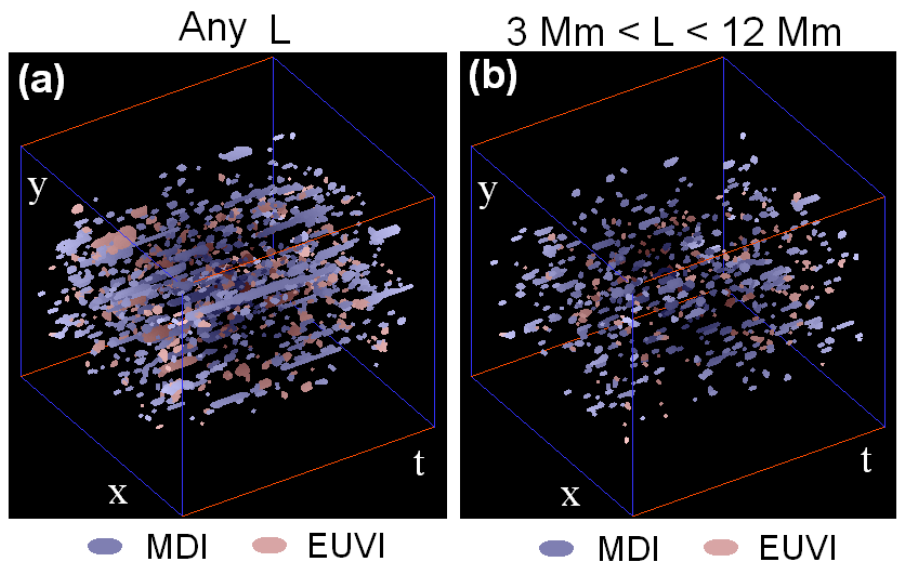} 
\caption{\label{fig4} Spatiotemporal plots showing about 1/4 of the photospheric (MDI) and coronal (EUVI) events detected respectively at $99 \%$ and $95 \%$ percentile levels. (a) No restriction on the liner size $L$ of the events is applied; (b) linear sizes are limitted by the interval of scales $L \in (3, 12)$ Mm used for evaluating scaling exponents. }
\end{center}
\end{figure*}

\begin{figure*}[htbp]
\includegraphics*[width=7.8 cm]{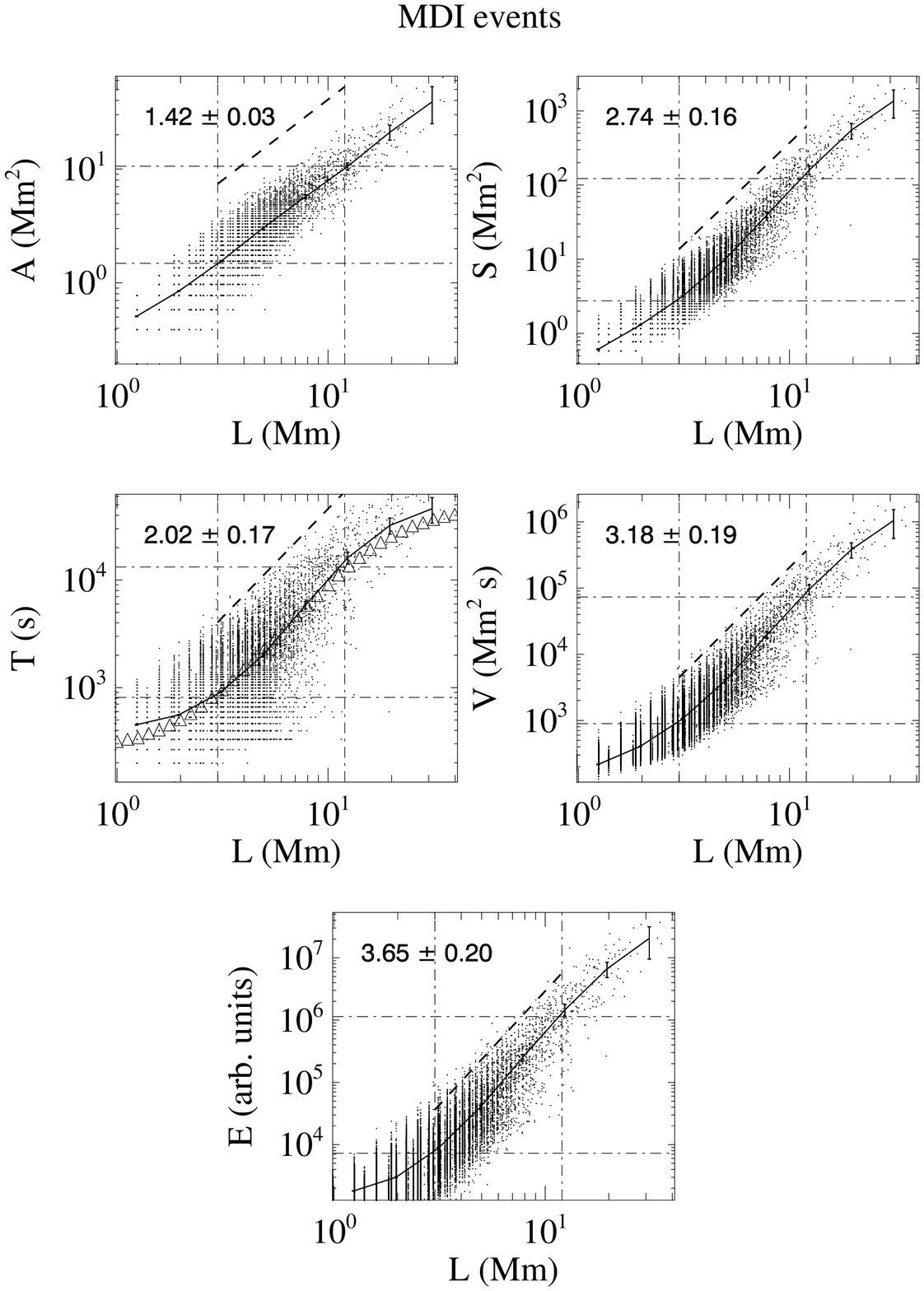} \hspace{.5cm} \includegraphics*[width=7.8 cm]{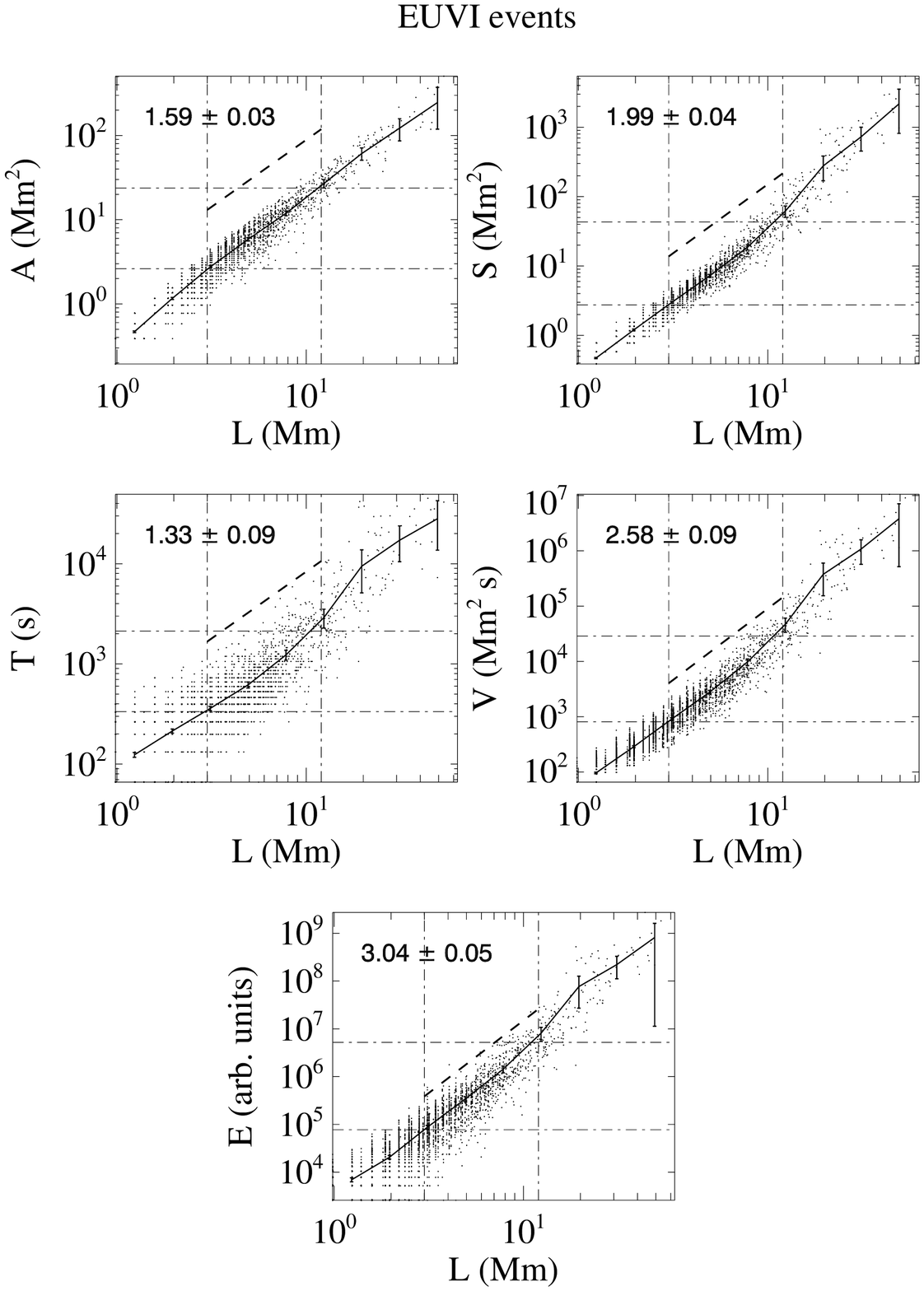} 
\caption{\label{fig5} Scatterplots of photospheric magnetic events (left) and coronal emission events (right) detected using the 95$\%$ percentile threshold. 
The solid lines show average trends obtained by rebinning the measured values of the scales size $L$ using 5 bins per decades. The error bars show three standard deviations around the mean value in each bin normalized by the square root of the number of points in this bin.}
\end{figure*}

\begin{figure*}[htbp]
\includegraphics*[width=7.8 cm]{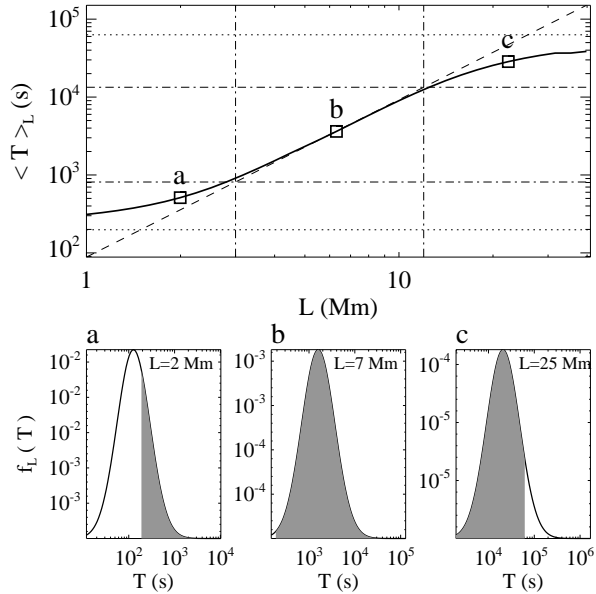}
\caption{\label{fig5_1} 
Top panel: The dependence of the conditional truncated mean value $\left \langle T \right \rangle_L$ on the linear scale $L$ (solid curve) showing the departure from the theoretical power-law dependence (dashed diagonal line) at smallest and largest scales due to an inadequate temporal resolution and limited duration of SOHO MDI data. 
Bottom panels: lognormal conditional probability density functions $f_L(T)$ corresponding to the points $a$, $b$ and $c$ on the top plot. White areas under the distributions are not sampled properly leading to the observed crossover effects. See text for details.} 

\end{figure*}

\begin{figure*}[htbp]
\includegraphics*[width=7.8 cm]{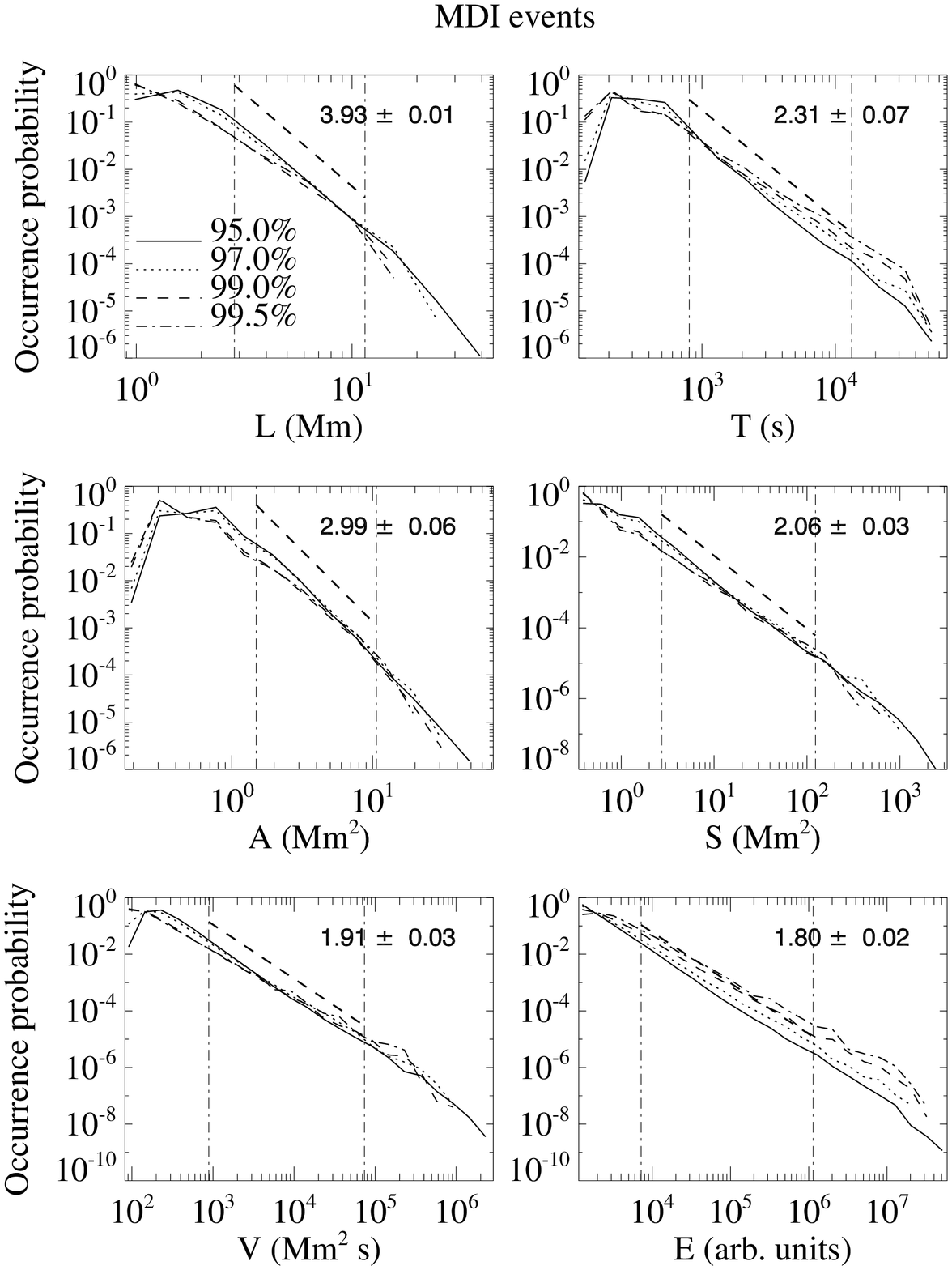} \hspace{.5cm} \includegraphics*[width=7.8 cm]{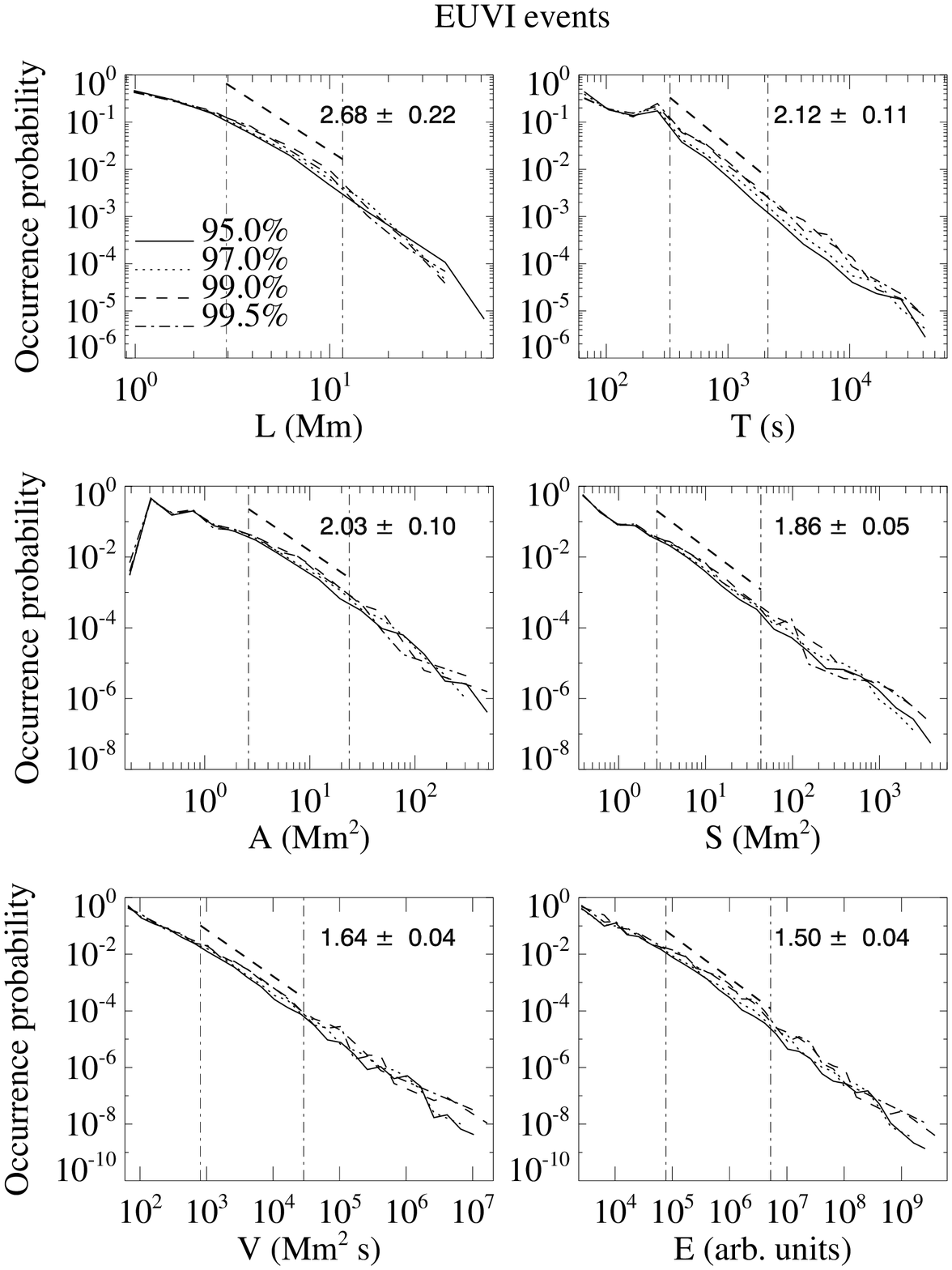} 
\caption{\label{fig6} Probability distributions of the photospheric magnetic events (left) and the coronal emission events (right) detected at four different percentile thresholds shown in Fig. \ref{fig3}. 
The power-law slopes and the standard errors of the slope provided in each panel refer to the $p = 95\%$ percentile threshold. }
\end{figure*}

\begin{figure*}[htbp]
\begin{center}
\includegraphics*[width=6.0 cm]{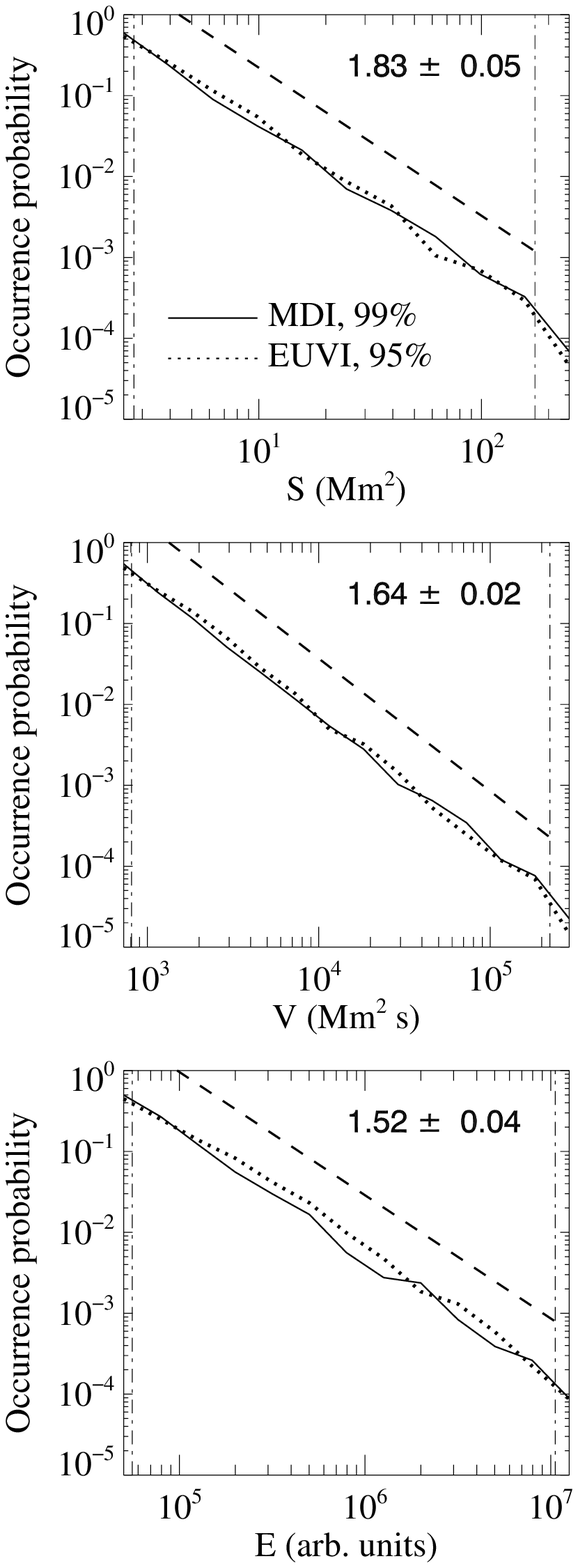} \hspace{.5cm} \includegraphics*[width=6.0 cm]{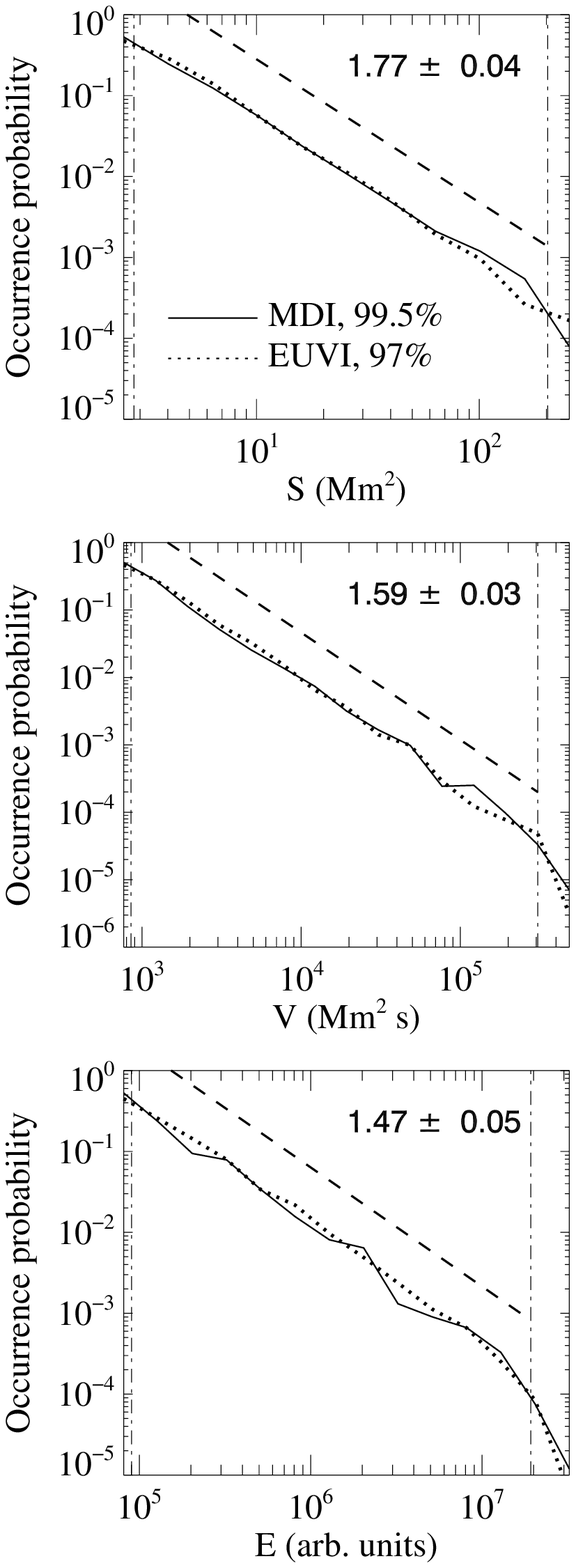} 

\caption{\label{fig7} Probability distributions of photospheric and coronal events obtained for two combinations of thresholds yielding comparable numbers of detected events in each data sets. The numerical values are the averaged log-log slopes of the MDI and EUVI distributions evaluated over a combined range of scales defined by a union of the MDI and EUVI scaling ranges bounded in Fig. \ref{fig6}(a) and (b) with vertical lines.} 
\end{center}
\end{figure*}

\begin{figure*}[htbp]
\begin{center}
\includegraphics*[width=7 cm]{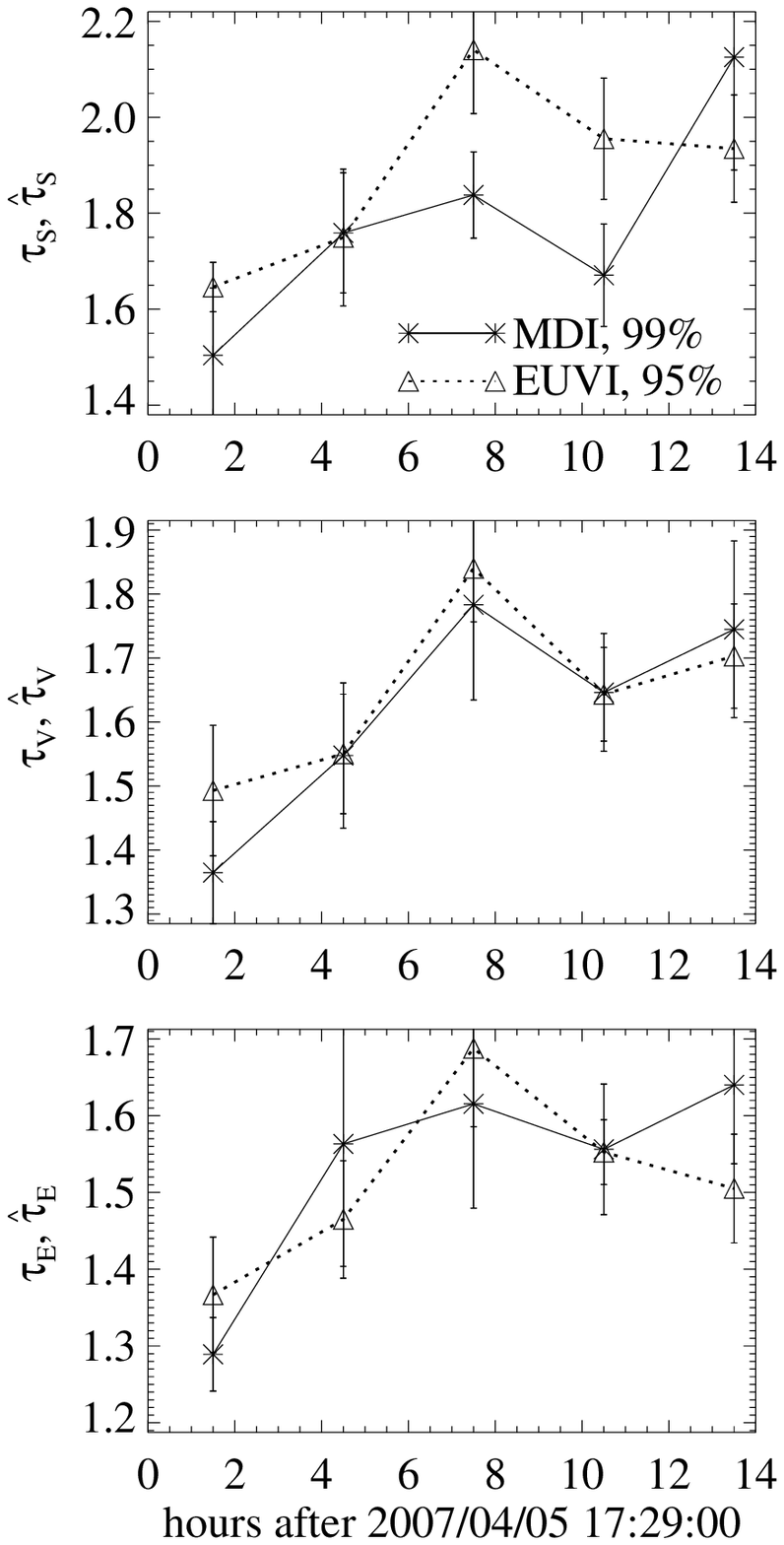} \hspace{.5cm} \includegraphics*[width=7 cm]{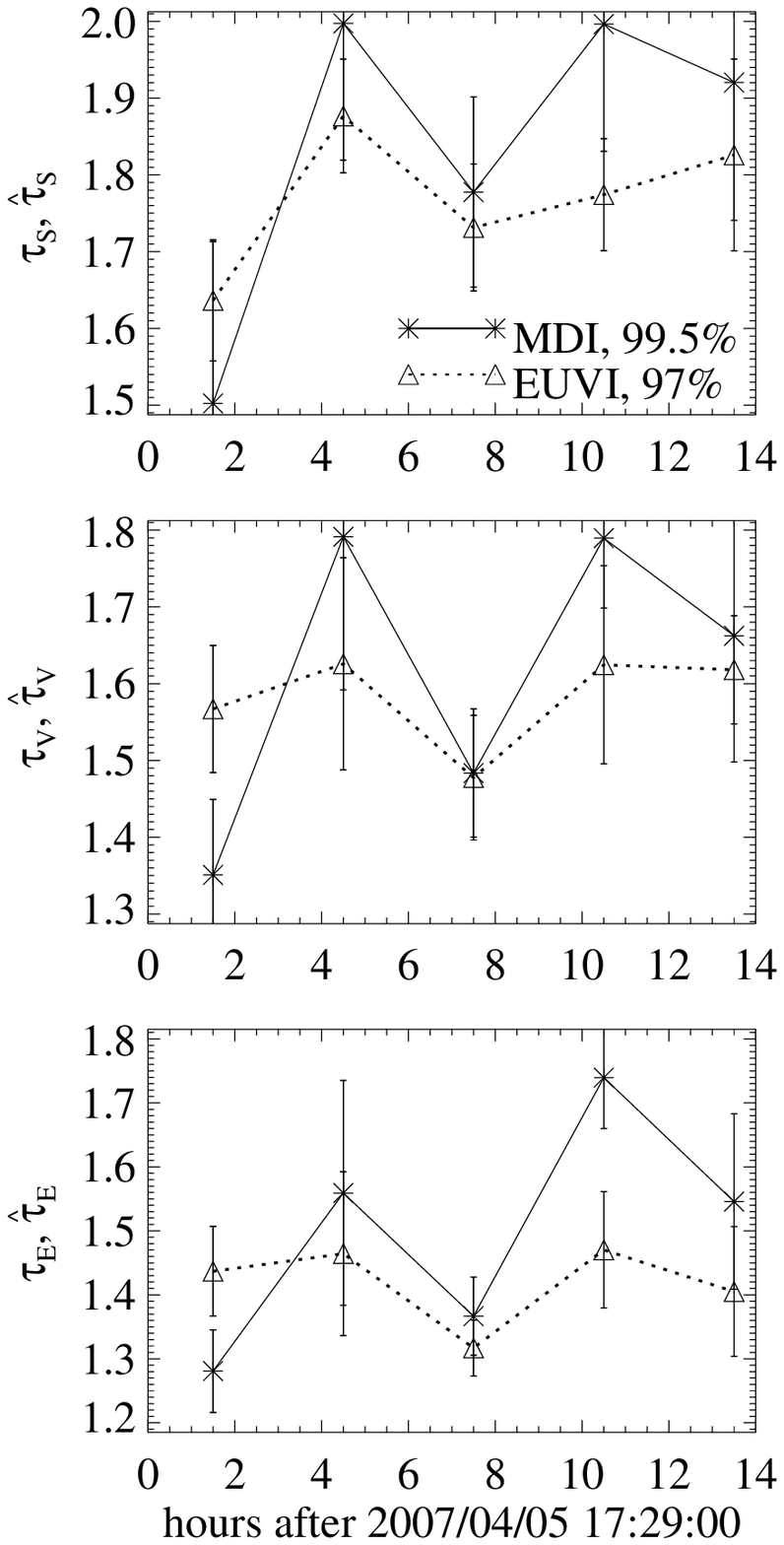}
\caption{\label{fig8} Temporal evolution of three distribution exponents of MDI and EUVI events computed within a 90-minute moving window. Vertical bars show the standard errors of the plotted values. }
\end{center}
\end{figure*}

\begin{figure*}[htbp]
\begin{center}
\includegraphics*[width=7 cm]{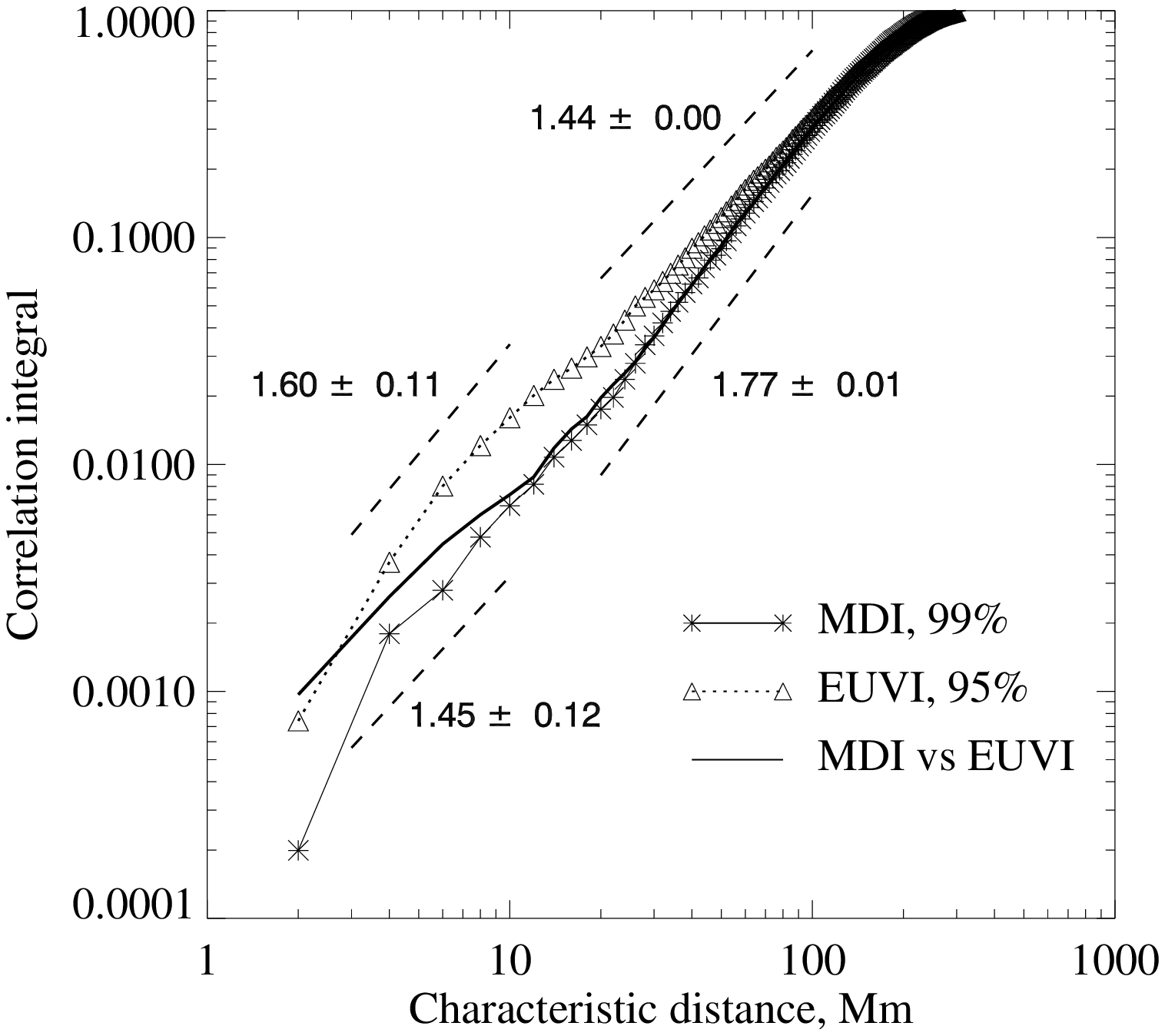} \hspace{.5cm} \includegraphics*[width=7 cm]{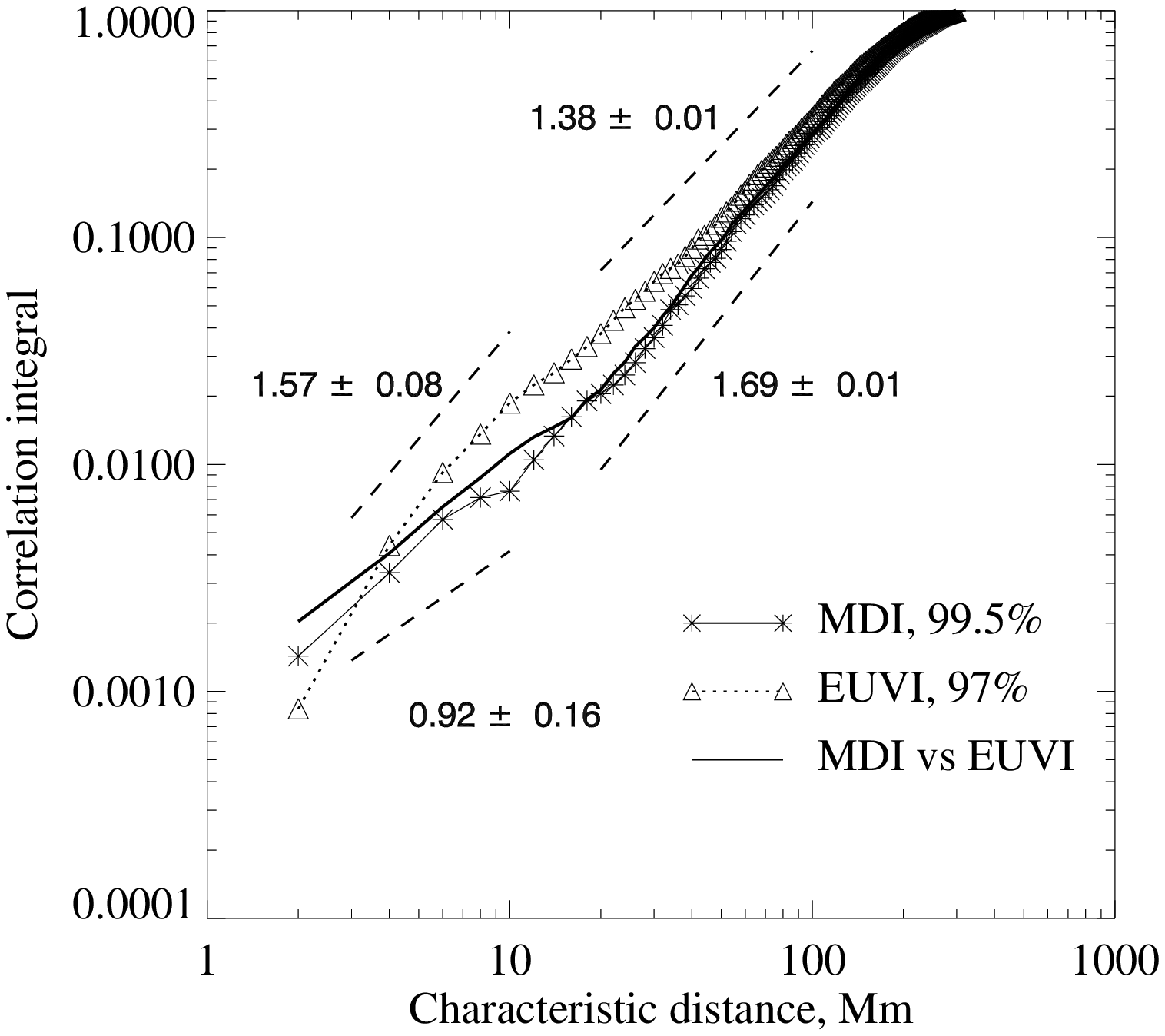}
\caption{\label{fig9} Auto- and cross-correlation integral analysis of the first occurrence locations of intermittent events in solar photosphere and corona, for two combinations of detection thresholds. Both solar regions exhibit significant spatial correlations across a wide range of distances. The values and standard errors of the CI exponents are provided for two subranges of scales: $3 - 10$ Mm and $20 - 100$ Mm. The log-log slopes of the cross-correlation integrals (solid lines) indicate significant positive correlation between coronal and photospheric events for distances $ < 15$ Mm.  }
\end{center}
\end{figure*}

\begin{figure*}[htbp]
\begin{center}
\includegraphics*[width=7 cm]{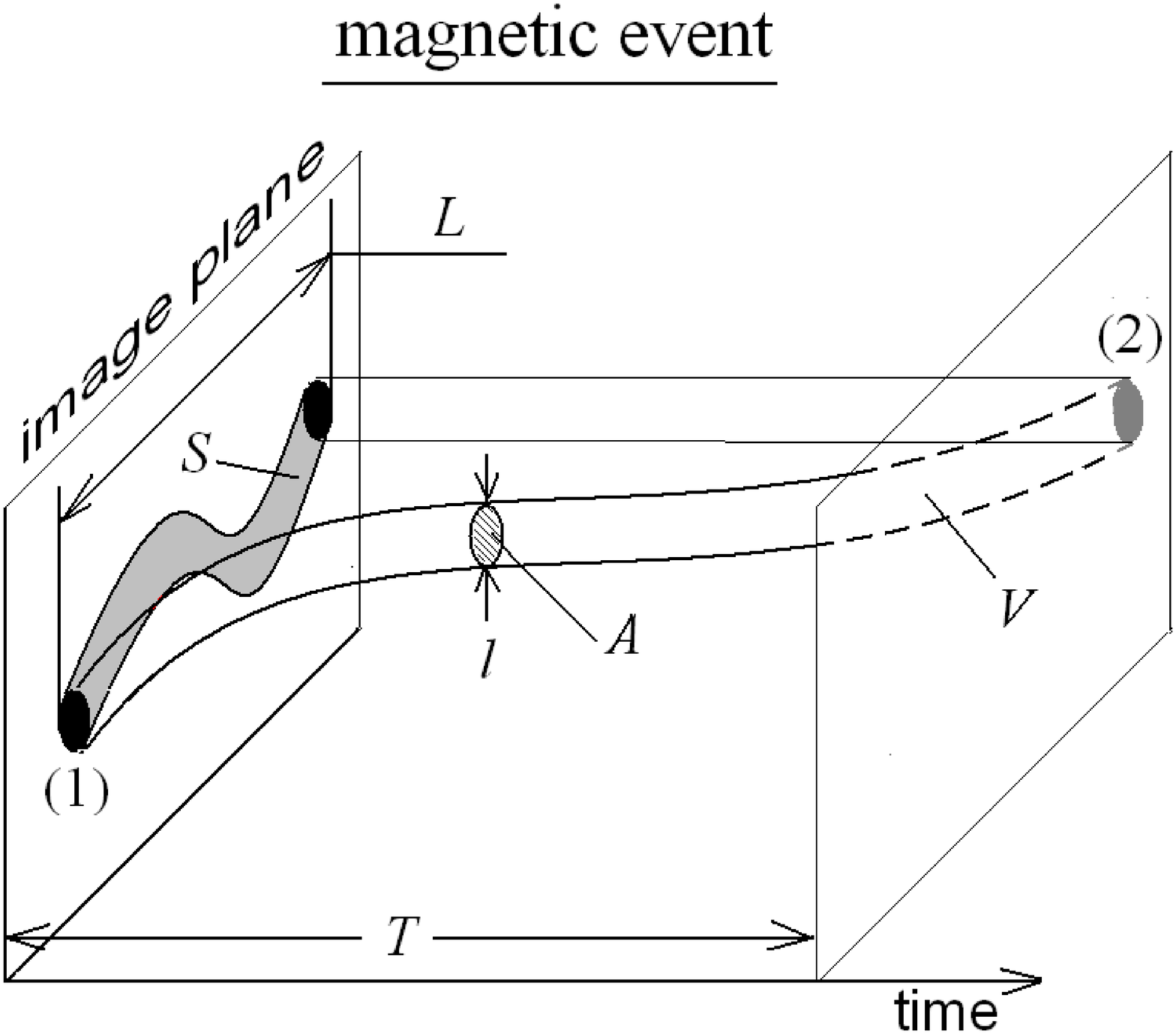} \hspace{.5cm} \includegraphics*[width=7 cm]{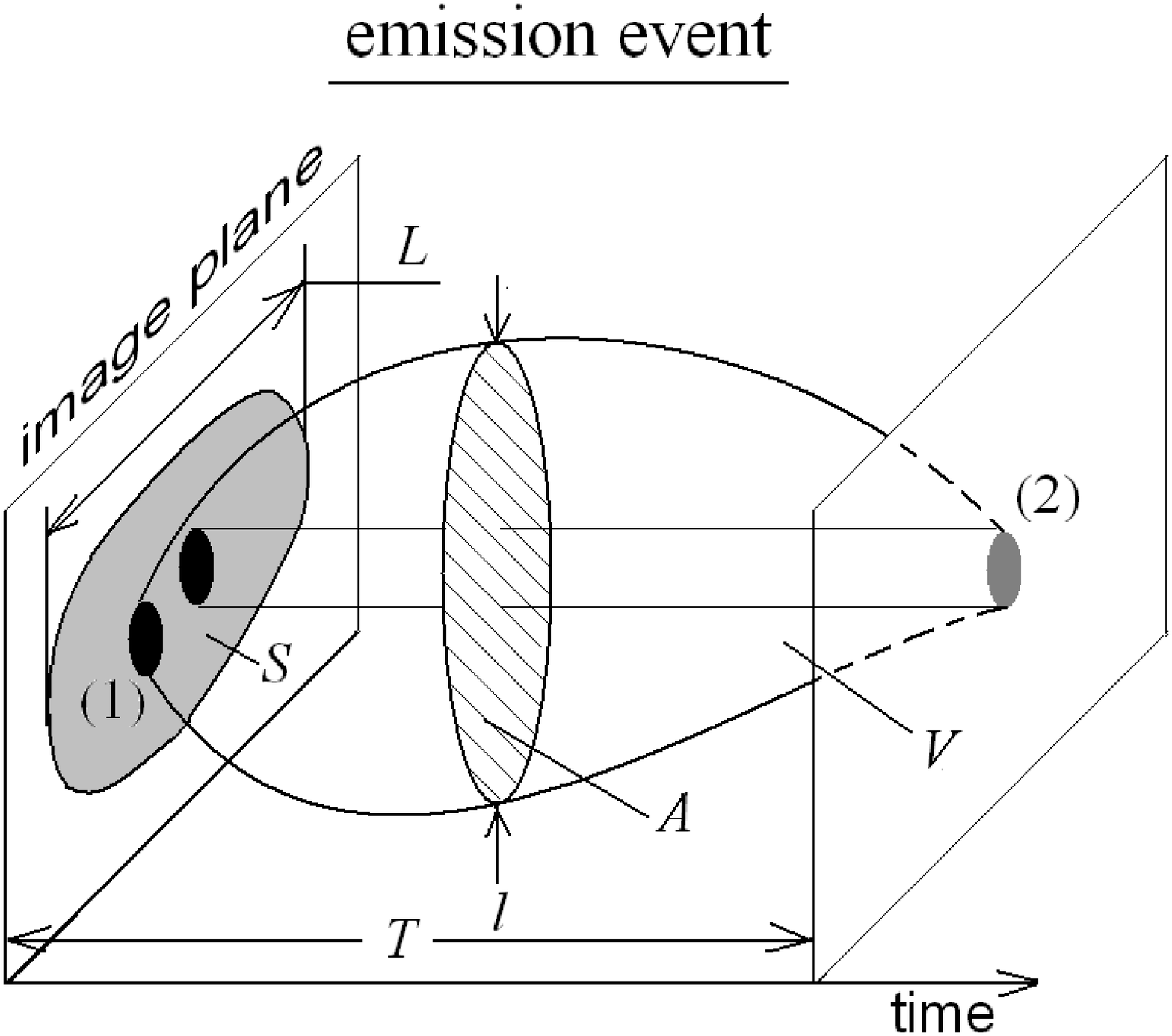}
\caption{\label{fig10} Schematic illustration of characteristic spatiotemporal geometries of magnetic and emission events reconstructed based on the performed scaling analysis. Labels (1) and (2) refer respectively to initial and final locations of the event; the shaded region is the projection of the event onto the image plane defining the swept area $S$.}
\end{center}
\end{figure*}

Due to the thresholding procedure which neglects most of the small-scale flares (Fig. \ref{fig3}), our statistics are likely to include relatively large events involving considerable plasma volumes, possibly on multiple interwoven coronal loops. For smaller events occurring inside individual loops, the photospheric length scales are obviously irrelevant because the magnetic field strength decreases by orders of magnitude from the photosphere to the corona, and the transverse scale expands accordingly. The multiscale dissipation pattern appearing at these scales must build up in the corona independently of the photospheric structuring, e.g. following the SOC or turbulence scenarios as we discussed in Section \ref{sec:mechanisms}.

In summary, we have shown that energy dissipation scales of coronal heating events can be controlled by turbulent photospheric convection. Our findings speak in favor of the coupling scenario proposed by Parker \citep{parker83, parker88} 
in which random photospheric shufflings generate marginally stable magnetic discontinuities at the coronal level. They are also consistent with the the Alfv\'{e}n wave heating mechanism \citep{davila87,ofman98,depontieu07} which can play a major part at shorter time scales. The large-scale processes such as those driven by super- and mesogranulation flows are likely to be providing a dominant portion of free magnetic energy dissipated in the corona, at least for the range of coronal emission intensities considered in our study. The question remains: is this energy sufficient to provide the bulk coronal heating necessary for the observed radiative loss rates?

\acknowledgments
We would like to thank J. Klimchuk, S. Antiochos, M. Aschwanden and M. Georgoulis for useful discussions and advice. The work of V.U. was supported by the NASA grant NNG11PL10A 670.002 through the CUA's Institute for Astrophysics and Computational Sciences. L.O. was supported by NASA grant NNX12AB34G.

\bibliographystyle{apj}




\clearpage

\begin{table*}
\begin{center}
\caption{\label{table1} Percentile thresholds used to detect intermittent events in SOHO MDI and STEREO EUVI image sets ($n$ - number of detected events).}
\begin{tabular}{lcc}
\hline
$p$ & MDI threshold & EUVI threshold    \\
\hline
95.0 $\%$  & 13.5 ($n=38242$) & 202 ($n=4124$) \\
97.0 $\%$  & 17.3 ($n=19114$) & 217 ($n=3005$) \\
99.0 $\%$  & 32.3 ($n=5912$) & 258 ($n=1269$)  \\
99.5 $\%$  & 42.0 ($n=3410$) & 292 ($n=686$)   \\
\hline
\end{tabular}
\end{center}
\end{table*}

\begin{table*}
\begin{center}
\caption{\label{table2} Scaling exponents of the photospheric and coronal events detected using four different values of the percentile threshold $p$.}
\begin{tabular}{l | ccc | ccc}
\hline
    & & MDI & & & EUVI & \\
\hline
$x$ & $D_x$ & $\tau_x$ & $\Delta_x$ & $\hat{D}_x$ & $\hat{\tau}_x$ & $\hat{\Delta}_x$\\
\hline
$p = 95 \%$ & & & & & & \\
$L$ &  1.00 $\pm$  0.00 &  3.93 $\pm$  0.01 &  2.94  & 1.00 $\pm$  0.00 &  2.68 $\pm$  0.22 &  1.66 \\
$T$ &  2.02 $\pm$  0.17 &  2.31 $\pm$  0.07 &  2.64 &  1.33 $\pm$  0.09 &  2.12 $\pm$  0.11 &  1.50 \\
$A$ &  1.42 $\pm$  0.03 &  2.99 $\pm$  0.06 &  2.82 &  1.59 $\pm$  0.03 &  2.03 $\pm$  0.10 &  1.64 \\
$S$ &  2.74 $\pm$  0.16 &  2.06 $\pm$  0.03 &  2.92 &  1.99 $\pm$  0.04 &  1.86 $\pm$  0.05 &  1.71 \\
$V$ &  3.18 $\pm$  0.19 &  1.92 $\pm$  0.02 &  2.93 &  2.58 $\pm$  0.09 &  1.64 $\pm$  0.04 &  1.66 \\ 
$E$ &  3.65 $\pm$  0.20 &  1.80 $\pm$  0.02 &  2.91 &  3.04 $\pm$  0.05 &  1.50 $\pm$  0.04 &  1.53 \\
\hline 

$p = 97 \%$ & & & & & &\\
$L$ &  1.00 $\pm$  0.00 &  3.67 $\pm$  0.19 &  2.67 &  1.00 $\pm$  0.00 &  2.45 $\pm$  0.15 &  1.44 \\
$T$ &  2.15 $\pm$  0.09 &  2.00 $\pm$  0.03 &  2.14$^*$ &  1.38 $\pm$  0.06 &  1.92 $\pm$  0.09 &  1.27$^*$ \\
$A$ &  1.40 $\pm$  0.02 &  2.80 $\pm$  0.02 &  2.52 &  1.60 $\pm$  0.03 &  1.82 $\pm$  0.02 &  1.31 \\
$S$ &  2.82 $\pm$  0.13 &  1.91 $\pm$  0.03 &  2.56 &  2.00 $\pm$  0.03 &  1.80 $\pm$  0.01 &  1.59$^*$ \\ 
$V$ &  3.32 $\pm$  0.14 &  1.75 $\pm$  0.03 &  2.49 &  2.61 $\pm$  0.04 &  1.60 $\pm$  0.04 &  1.57 \\
$E$ &  3.77 $\pm$  0.12 &  1.65 $\pm$  0.03 &  2.46 &  3.07 $\pm$  0.03 &  1.43 $\pm$  0.04 &  1.31 \\
\hline 

$p = 99 \%$ & & & & & &\\
$L$ &  1.00 $\pm$  0.00 &  3.08 $\pm$  0.03 &  2.12 &  1.00 $\pm$  0.00 &  2.11 $\pm$  0.17 &  1.08 \\
$T$ &  2.21 $\pm$  0.06 &  1.84 $\pm$  0.04 &  1.84$^*$ &  1.34 $\pm$  0.14 &  2.04 $\pm$  0.11 &  1.39$^*$ \\ 
$A$ &  1.34 $\pm$  0.04 &  2.53 $\pm$  0.03 &  2.06 &  1.60 $\pm$  0.09 &  1.77 $\pm$  0.13 &  1.23$^*$ \\
$S$ &  2.75 $\pm$  0.09 &  1.79 $\pm$  0.04 &  2.18 &  1.99 $\pm$  0.01 &  1.75 $\pm$  0.07 &  1.50$^*$ \\
$V$ &  3.39 $\pm$  0.10 &  1.63 $\pm$  0.04 &  2.15 &  2.63 $\pm$  0.03 &  1.56 $\pm$  0.06 &  1.47$^*$ \\
$E$ &  3.78 $\pm$  0.10 &  1.52 $\pm$  0.04 &  1.96 &  3.01 $\pm$  0.02 &  1.47 $\pm$  0.06 &  1.42$^*$ \\
\hline 

$p = 99.5 \%$ & & & & & &\\
$L$ &  1.00 $\pm$  0.00 &  3.22 $\pm$  0.35 &  2.30 &  1.00 $\pm$  0.00 &  2.53 $\pm$  0.01 &  1.54 \\
$T$ &  2.14 $\pm$  0.08 &  1.74 $\pm$  0.03 &  1.58$^*$ &  1.28 $\pm$  0.04 &  2.32 $\pm$  0.14 &  1.69 \\
$A$ &  1.43 $\pm$  0.01 &  2.50 $\pm$  0.10 &  2.14 &  1.62 $\pm$  0.11 &  1.78 $\pm$  0.15 &  1.26$^*$ \\
$S$ &  2.81 $\pm$  0.19 &  1.68 $\pm$  0.03 &  1.90$^*$ &  1.95 $\pm$  0.03 &  1.86 $\pm$  0.10 &  1.68 \\
$V$ &  3.46 $\pm$  0.21 &  1.49 $\pm$  0.07 &  1.68$^*$ &  2.60 $\pm$  0.00 &  1.58 $\pm$  0.07 &  1.51 \\
$E$ &  3.82 $\pm$  0.22 &  1.44 $\pm$  0.04 &  1.68$^*$ &  2.97 $\pm$  0.06 &  1.58 $\pm$  0.10 &  1.71 \\ 
\hline 

\end{tabular}
\end{center}
\end{table*}

\begin{table*}
\begin{center}
\caption{\label{table3} Average values of scaling exponents obtained by combining $p = 99\%$ and $p = 99.5\%$ data for the photospheric events with $p = 95\%$ and $p = 97\%$ data for the coronal events as explained in the text. The distribution exponents of time-integrated event measures are highlighted with bold font. }
\begin{tabular}{lcccc | c}
\hline
\\
$x$ & $D_x$ & $\hat{D}_x$ & $\tau_x$ & $\hat{\tau}_x$ & $\hat{\tau}_x$, predicted \\
\\
\hline
\\
$L$ &  1.00 $\pm$  0.00 &  1.00 $\pm$  0.00 &  3.15 $\pm$  0.18 &  2.57 $\pm$  0.13 & 2.55 \\ 
\\
$T$ &  2.18 $\pm$  0.05 &  1.36 $\pm$  0.05 &  1.79 $\pm$  0.03 &  2.02 $\pm$  0.07 & 1.91 \\ 
\\
$A$ &  1.39 $\pm$  0.02 &  1.60 $\pm$  0.02 &  2.52 $\pm$  0.05 &  1.93 $\pm$  0.06 & 1.95 \\ 
\\
$S$ &  2.78 $\pm$  0.11 &  2.00 $\pm$  0.03 &  \textbf{1.74 $\pm$  0.03} &  \textbf{1.83 $\pm$  0.03} & \textbf{1.74} \\ 
\\
$V$ &  3.43 $\pm$  0.12 &  2.60 $\pm$  0.05 &  \textbf{1.56 $\pm$  0.04} &  \textbf{1.62 $\pm$  0.03} & \textbf{1.56} \\ 
\\
$E$ &  3.80 $\pm$  0.12 &  3.06 $\pm$  0.03 &  \textbf{1.48 $\pm$  0.03} &  \textbf{1.47 $\pm$  0.03} & \textbf{1.42} \\ 
\\
\hline

\end{tabular}
\end{center}
\end{table*}

\clearpage

\end{document}